\documentstyle[12pt,aasms4]{article}
\slugcomment{}
\lefthead{Enya et al.}
\righthead{$JHK'$ variability of AGNs}

\begin{document}

\title{$JHK'$ Imaging Photometry of Seyfert 1 AGNs and Quasars III: \\
        Variability of Radio Quiet and Radio Loud AGNs}

\author{Keigo Enya\altaffilmark{1}, Yuzuru Yoshii\altaffilmark{1,4},
   Yukiyasu Kobayashi\altaffilmark{2}, Takeo Minezaki\altaffilmark{1}, 
   Masahiro Suganuma\altaffilmark{3}, Hiroyuki Tomita\altaffilmark{3}
   and Bruce A. Peterson\altaffilmark{5}}

\vspace{20mm}

\affil{$^1$ Institute of Astronomy, School of Science,
              University of Tokyo, Osawa 2-21-1,  
              Mitaka, Tokyo 181-8588, Japan}
\affil{$^2$ National Astronomical Observatory,
              Osawa 2-21-1, Mitaka, Tokyo 181-8588, Japan}
\affil{$^3$ Department of Astronomy, University of Tokyo,
             Hongo 7-3-1, Bunkyo-ku, Tokyo  113-0033, Japan}
\affil{$^4$ Research Center for the Early Universe (RESCEU),
          School of Science, University of Tokyo,
             Hongo 7-3-1, Bunkyo-ku, Tokyo 113-0033, Japan}
\affil{$^5$ Research School of Astronomy and Astrophysics, The
         Australian National University, Weston Creek, ACT 2611,
         Australia}


\begin{abstract}

Variability of 226 AGNs in the near-infrared $J$, $H$, and $K'$ bands 
is analyzed and discussed.  An ensemble average for measured 
variabilities was obtained for various samples of the AGNs divided by 
absolute $B$-magnitude $M_B$, redshift $z$, and radio strength.  
All the samples in the $J$, $H$, and $K'$ bands are found to give  
significant ensemble variability, but no significant wavelength dependence 
is found.  
The ensemble variability in the entire sample combining the $J$, $H$, 
and $K'$ samples is $\Delta m\approx 0.22$ mag, while $\Delta m\approx 
0.18$ mag for the radio-quiet AGNs and $\Delta m\approx 0.26$ mag 
for radio-loud AGNs.  The ensemble variability for the radio-quiet 
AGNs shows no significant $M_B$-dependence, while showing 
positive $M_B$-dependence for the radio-loud AGNs.
In any samples the measured variability shows positive correlation 
among different passbands, with the correlation coefficients of 
$r_{JH}$, $r_{HK'}$, and $r_{JK'}$ ranging from 0.6 to 0.9.  
For radio-quiet AGNs, the coefficient $r_{HK'}$ in a redshift range of 
$0.1<z<0.3$ is significantly higher than $r_{JH}$ or $r_{JK'}$. The 
coefficient for the radio-loud AGNs with $0.6<z<1.0$ is as high as 0.95, 
irrespective of the passband.  However, for the radio-quiet AGNs 
with $z>0.3$ and radio-loud AGNs with $z<0.3$, we cannot confirm such 
strong correlation among different passbands.
All the features of near-infrared variability for the radio-quiet 
AGNs are  consistent with a simple dust reverberation model of  the 
central regions of AGNs.  However, the features for the radio-loud AGNs 
are not fully explained by such a model, and a non-thermal variable 
component is suggested as a viable candidate for  causing their large and 
fast variability in the near-infrared region.

\end{abstract}


\section{Introduction}

Active galactic nuclei (AGNs) at cosmological distance emit enormous 
amount of 
energy from their central region which is compact and  
spacially unresolved.  Observations of AGN variability are important
for the understanding of  the physical mechanisms of energy emission 
by the central engine.

Monitoring observations of many AGNs in the optical region have 
been made by various authors, and  the relations among AGN optical 
variability, luminosity, redshift, time scale and so on have been 
derived.  For example, a negative correlation between variability 
amplitude and luminosity was found,  and the possibility of explaining
the emission and variability of AGN by a sub-unit model has been 
discussed (e.g., Cristiani et al. 1996; Hook et al. 1994).  
The wavelength dependence 
of variability was discussed by various authors 
(e.g., Cristiani et al. 1997; Winkler 1997; Winkler et al. 1992).  
Multi-color monitoring observations of AGNs are useful to distinguish 
the dependence of valiability on wavelength and redshift.  Optical 
variability was compared with ultraviolet variability (Clemente et al. 
1996), and it was concluded that the variability was larger in the 
shorter wavelength UV region than in the optical region.

Further understanding can obviously be made  by combining the 
near-infrared (NIR) data with the UV/optical data.  Barvanis (1992) 
analyzed the optical and NIR light curves of Fairall 9 (Cravel, 
Wamsterker \& Glass 1989), and explained the delay of NIR variability, 
in terms of a dust reververation model, 
in which  thermal re-radiation comes from a hot 
dust torus illuminated by the central emission engine.  Neugebauer et al. 
(1988) monitored 108 PG quasars in the $J$, $H$, $K$, $L$ and 10$\mu$m 
bands for about 20 years.  It was suggested that the detection rate  
of variability is smaller in the NIR than in the optical, 
if the same level of accuracy is required.

We present the new data of $J$, $H$, and $K'$ variabilities for 226 
AGNs. The sample was divided into groups by radio strength, absolute 
$B$-magnitude, and redshift, 
and the ensemble variability for each group was measured.

\section{Data}

Our analysis in this paper was made by using the NIR images of AGNs 
obtained through the procedure of image reduction discussed in Paper I.  
Differential photometry was employed to measure their NIR variability.
The achieved accuracy was significantly higher than the accuracy from 
an alternative method of standards-based photometry, as described in 
Paper II.  Only the measurements made with more than two reference objects 
and having an accuracy higher than 0.1 mag are used in this paper.

The AGNs in our sample were selected from various versions of the Quasars 
and Active Galactic Nuclei catalog (VV catalog, Veron-Cetty and Veron 
1993, 1996, 1998).  AGNs were selected with  consideration of their use 
in the MAGNUM Project (Kobayashi et al. 1998a, 1998b).  
The distribution of declination and right 
ascention for all AGNs in the sample is shown in Fig. 1 of Paper I.  
The distribution of absolute $B$-magnitude and redshift is shown in 
Fig. 2 of Paper I.

All observations were made with the 1.3m infrared telescope at the  
Institute of Space and Astronautical Science (ISAS), Japan, equipped 
with the NIR PICNIC camera (Kobayashi et al. 1994).  The AGNs and reference 
objects were imaged in the $J$, $H$ and $K'$ bands with  
the telescope stepped in a raster pattern.  
Two photometric standard stars with different 
elevations were observed three times each night (for details see \S 2 of
Paper I).

The PICRED software, that was developed for the PICNIC camera, was
used to reduce the images.  During our observations this software 
was optimized for  AGN and quasar images, to achieve  fully automated 
reductions (for details see \S 3 of Paper I).

\section{Analysis and Discussion}

\subsection{ensemble variability}

Since each AGN was observed on two  different nights separated 
by a year or more,  the variability thus obtained does not necessarily 
reflect the characteristic amplitude of intrinsic variability.  
Therefore, 
we discuss the ensemble variability of AGNs which reflects the dispersion 
of their individual variabilities.  However, the 
standard deviation of such data is not a good parameter in estimating 
the AGN variability, because not only the intrinsic AGN variability but 
also measurement errors broaden the distribution.  

We here introduce the ensemble variability, after excluding the 
contribution of measurement errors, as
\begin{equation}
\Delta m=\sqrt{ \frac{\sum_i^N{\Delta m_i^2} - \sum_i^N{\sigma_i^2}}{N} 
}\label{eq_V} \;\;\; ,
\end{equation}
and its error given by
\begin{equation}
\sigma_{\Delta m}=\frac{1}{2\Delta m} \sqrt{ \frac{\left(\sum_{i}^{N} 
\Delta m_i^2 - \sum \sigma_i^2\right)^2}{N^3}+
\frac{\sum_i^N\left(4\Delta m_i^2 \sigma_i^2 -
2\sigma_i^4\right)}{N^2} }\label{eq_V_err} \;\;\;.
\end{equation}
Derivation of $\Delta m$ and $\sigma_{\Delta m}$ is described in 
Appendix \ref{apnd_V}.  In the remainder of this paper these quantities 
of $\Delta m$ and $\sigma_{\Delta m}$ are  used to discuss the variability 
of the  AGNs in our sample.

\subsection{The relation between the variability and AGN 
character}\label{sub_V}

We examine whether the variability is correlated with the parameters 
such as radio strength, rest-frame time interval of observations, 
absolute $B$-magnitude, redshift, Seyfert type, and NIR colors. 
First, the sample was divided by each parameter into two groups, ``$a$'' 
and ``$b$'', 
at a point where there appeared to be a boundary
on either side of which 
the data are separated.  Then, the ensemble variabilities for these two 
groups were compared with each other.  

The AGNs in our sample are distinguished by the radio strength  or 
the ratio of radio 6 cm flux relative to optical $V$-band flux 
$f_{\nu}({\rm 6cm})/f_{\nu}(V)$. The AGNs with 
$f_{\nu}({\rm 6cm})/f_{\nu}(V)<10$ were classified into the radio-quiet 
group, and others with $f_{\nu}({\rm 6cm})/f_{\nu}(V)>100$ into the 
radio-loud group.  The AGNs, observed twice separated by time interval 
$\Delta t_{\rm obs}$, were distinguished by the rest-frame time interval 
$\Delta t_{\rm rest}\equiv\Delta t_{\rm obs}/(1+z)$. The AGNs with 100 days 
$<\Delta t_{\rm rest}<$400 days were classified into the short-$\Delta 
t_{\rm rest}$ group, and others with 400 days $<\Delta t_{\rm rest}<$ 
800 days into the long-$\Delta t_{\rm rest}$ group.  The boundary 
at 400 days reflects the period in which the AGNs were observed.  Our 
observational runs consist of three periods (January 1996$-$April 1996, 
November 1996$-$February 1997, December 1997$-$April 1998).  The AGNs 
observed in the first and third periods are mainly of long interval, 
and those in the second and third periods are of short interval. The AGNs 
with $M_B<-23.5$ were classified into the bright group, and those with 
$M_B>-23.5$ into the faint group.  The AGNs with $z<0.3$ are classified 
into the low-$z$ group, and those with $z>0.3$ into the high-$z$
group.  The AGNs with Seyfert 1, 1.2, and 1.5 were classified into the 
early-type Seyfert group, and others with 1.8, 1.9, and 2 into the 
late-type Seyfert group. The AGNs with $J-H<0.8$ and those with $H-K'<0.8$ 
were classified into the blue group, and others with $J-H>0.8$ and 
$H-K'>0.8$ into the red group.

The ensemble variabilities for the ``$a$'' and ``$b$'' groups are 
derived for each of radio strength, rest-frame time interval of 
observations, absolute $B$-magnitude, redshift, Seyfert type, 
and NIR colors.  Table 1 shows the ratio 
$\Delta m(``a")/\Delta m(``b")$ 
for each of the above parameters. The last column of this table 
represents the average ratio taken over  the $J$, $H$, and $K'$ 
bands.  The average ratio between the radio-loud and radio-quiet 
groups is 1.46, which is the largest.  The average 
ratio for the long- and short-$\Delta t_{\rm rest}$ groups is 1.24, 
while it is 1.20 for the bright- and faint-$M_B$ groups.  
The average ratios for other quantities are much closer to unity.

Top panel of Fig. 1 shows that the ensemble 
variability in the $J$, $H$, or $K'$ band is $\Delta m\approx 
0.2$, estimated for the entire sample, while 
$\Delta m\approx 0.25-0.3$ and $0.18-0.2$ for the radio-quiet 
and radio-loud groups, respectively. 
We see no significant wavelength-dependence of $\Delta m$.

We divide the radio-quiet or radio-loud group furthermore into two 
subgroups of short and long $\Delta t_{\rm rest}$, and estimate the 
ensemble variability for respective subgroups.  Bottom panels of Fig. 1 
show that the ensemble variability in the $J$, $H$, or $K'$ band is 
$\Delta m\approx 0.2$ (long $\Delta t_{\rm rest}$) and 0.15 
(short $\Delta t_{\rm rest}$) for the radio-quiet sample, while 
$\Delta m\approx 0.28-32$ (long $\Delta t_{\rm rest}$) 
and $0.22-0.28$ (short $\Delta t_{\rm rest}$) for the radio-loud 
sample.  

We furthermore divide the short- or long-$\Delta t_{\rm rest}$ 
subgroup by $M_B$ or $z$. In this way, the ensemble variabilities
of the radio-quiet AGNs in respective subgroups are shown in Fig. 2.  
The similar results for the radio-loud AGNs are shown in Fig. 3.

\subsubsection{statistical  test and estimation of 
    the $\lambda$ dependence of the ensemble variability}

A statistical test on the wavelength dependence of $\Delta m_{\lambda}$ 
was done by applying the $\chi^2$ method to the result.  
Since such dependence was not found to be  significant, 
it is reasonable to adopt a two-parameter function of 
$\Delta m_{\lambda}(a_1, a_2)=a_1 e^{a_2\lambda}$ and search for the 
solution near $a_2\approx 0$ in minimizing 
$\chi^2=\sum_{\lambda=J,H,K'} (\Delta m_\lambda-
\Delta m_\lambda(a_1, a_2))^2/\sigma_{\Delta m_\lambda}^2$.
Table 2 shows the optimized values of $a_1=0.205$
and $a_2=0.029$ for the entire sample, leading to 
$\Delta m_{K'}/\Delta m_J=1.03$, otherwise 
$\Delta m_{K'}/\Delta m_J=0.83-1.26$ (95\% C.L.) and $0.79-1.32$ 
(99\% C.L.).  For the radio-quiet sample, the optimized values of 
$a_1=0.205$ and $a_2=-0.065$ give $\Delta m_{K'}/\Delta m_J=0.94$.   
A similar result holds if the radio-quiet sample is further divided by 
$\Delta t_{\rm rest}$. For other groups, the C.L. range becomes wider, 
so that the ability of rejecting the hypothesis of 
no wavelength-dependence by the test remarkably decreases.

\subsubsection{statistical  dependence test of the ensemble
    variability on characteristic parameters}

In this section, using the ensemble variability $\Delta m$ averaged 
over the $J$, $H$, and $K'$ bands, we examin whether $\Delta m$ 
depends on radio strength, $\Delta t_{\rm rest}$, $M_B$, and $z$. 
In this case, with a simple one-parameter function of 
$\Delta m(a_1)=a_1$, it is reduced to 
$\chi^2=\sum_j(\Delta m_j-\Delta m(a_1))^2/\sigma_{\Delta 
m_j}^2$.  Table 3 shows the optimized value of 
$a_1$ together with $P$ which represents the reliability of rejecting 
the hypothesis that $\Delta m$ does not depend on the parameter in 
question.

It is understood from this table that the statistical equivalence of 
$\Delta m$ between the radio-quiet and radio-loud samples is rejected 
by a level of $P\ge 99.9\%$.  For the radio-quiet sample further 
divided by $\Delta t_{\rm rest}$, the statistical equivalence of $\Delta 
m$ between the short and long $\Delta t_{\rm rest}$ is also rejected 
by a level of $P\ge 99.9\%$.  For the radio-loud sample, however, such 
statistical equivalence is rejected only by $P=71.1\%$.

For the radio-quiet sample with short $\Delta t_{\rm rest}$, the test 
for the $M_B$-dependence gives $P=75.0\%$ by which it is difficult to 
conclude with certainty that $\Delta m$ depends on $M_B$.  The same 
test for $z$-dependence gives $P=95.5\%$, indicating a rather strong  
$z$-dependence of $\Delta m$.  For the radio-loud sample with short 
$\Delta t_{\rm rest}$, both $M_B$-dependence and $z$-dependence are 
highly significant.  We note that such strong $M_B$-dependence is in 
clear contrast with the result for the radio-quiet sample, and such 
strong $z$-dependence is similar to the result for the radio-quiet 
sample. 

\subsubsection{discussion of the parameter dependence 
     of the ensemble variability}

Figure 4 shows the relations among various parameters for the 
radio-quiet sample. In the left column, from top to bottom, are 
shown the values of $\Delta m$, $\Delta t_{\rm rest}$, and $M_B$, 
estimated in four $z$-bins between $z=0$ and 1.  Similarly, 
in the right column, from top to bottom, are shown the values of 
$\Delta m$, $\Delta t_{\rm rest}$, and $z$, estimated in four $M_B$-bins 
between $M_B=-32$ and $-18$. In this figure, open and filled squares 
represent the samples with short and long $\Delta t_{\rm rest}$, 
respectively. 

We see from the left column that $\Delta m$, $\Delta t_{\rm rest}$ 
and $M_B$ decrease with increasing $z$, except for the case of $\Delta 
m$ in the lowest-$z$ bin.  These trends with $z$ are equivalently 
converted to the trends with $M_B$, as shown in the right column, by 
using the monotonical $z$ versus $M_B$ relation.   We  caution 
that the $z$-dependence of $\Delta t_{\rm rest}$ may only be apparent, 
arising from the cosmological time delay $\Delta t_{\rm rest}=\Delta 
t_{\rm obs}/(1+z)$ applied to our sample which has  a rather limited range 
of $\Delta t_{\rm obs}$. Consequently, any trends with $\Delta t_{\rm 
rest}$ may also be apparent.

The ensemble NIR variability of radio-quiet AGNs is as small as $\Delta 
m\le 0.25$, showing little $M_B$-dependence in this work.  We  note 
that faint AGNs, mostly at low $z$, are contaminated by a host galaxy 
component, which is indicated by our multi-aperture color analysis 
(Paper I; see also Kotilainen \& Ward 1994).  Since the host galaxy 
component is stellar and 
does not vary on a time scale of  years, 
such contamination has the  systematic 
effect of weakening the AGN variability.  Therefore, $\Delta m$, 
after correction  for this effect, would still have little or negative 
correlation with absolute $B$-luminosity, depending on the degree of 
contamination within chosen aperture size.  

The above $M_B$-dependence of $\Delta m$ is expected from the model of 
dust reverberation in which brighter AGNs have a larger dust torus. That 
is, a variation in the  UV/optical light emitted from the central engine is
absorbed in  more extended region of dust from which the 
NIR radiation is emitted and the spread in arrival times of the NIR 
variation from this extended region produces a variation with smaller
amplitude.  
Therefore, regardless of the real emission 
mechanism of central source, the NIR variability would show only small 
correlation with $M_B$, as observed.


Next, we consider the radio-loud AGNs.  Figure 5 
shows their relations among $\Delta m$, $\Delta t_{\rm rest}$, $M_B$, 
and $z$, in a similar way as in Fig. 4.  
We divide the radio-loud sample into the short and long $\Delta t_{\rm 
rest}$.  However, because of the lack of enough data, the values of 
$\Delta m$ and $\Delta t_{\rm rest}$ are not estimated as a function 
of $z$ or $M_B$, for the case of long $\Delta t_{\rm rest}$.

We notice that $\Delta m$ for the radio-loud AGNs, in the case 
of short $\Delta t_{\rm rest}$, strongly increases with increasing $z$ 
or with increasing absolute $B$-luminosity, 
opposite to that for the radio-quiet 
AGNs.  If we assume that such positive correlation is only apparent, 
wishing to explain it in terms of the effect of contamination of host 
galaxy component, we have to invoke an extremely different
contribution from the  
host galaxies between radio-loud and radio-quiet AGNs, which is difficult 
to justify.  Therefore, it is more reasonable to conclude that the 
different $M_B$-dependence of $\Delta m$ reflects the different emission 
and variability mechanisms  between radio-loud and radio-quiet AGNs. 

\subsection{The correlation of variability in different 
passbands}\label{sub_band_soukan}

In this section the correlation among the variabilities in the $J$, $H$, 
and $K'$ bands is discussed using only the data of estimated variabilities 
with more than two reference stars and an  accuracy better than 0.1 mag.  
The results for the radio-quiet and radio-loud AGNs are shown in Figs. 
6 and 7, respectively, where $\Delta m_H$ is plotted against $\Delta m_J$, 
$\Delta m_K'$ against $\Delta m_H$, and $\Delta m_K'$ against $\Delta m_J$.  
Open and filled symbols correspond to long and short $\Delta t_{\rm rest}$, 
respectively. 

The data distribute, more or less, along the diagonal running from lower 
left to upper right through the origin, which indicates that AGNs, becoming 
brighter in one band, become brighter in the other band, and  vice versa.  
It is 
seen from each panel that more data are plotted in the lower left region than 
in the upper right region.  However, this is not real because AGNs, becoming 
fainter, are likely to be either undetected or rejected by our accuracy 
requirement.


The correlation coefficient of variabilities in two different bands, 
say $J$ and $H$, is defined as
\begin{equation}
r_{JH}=\frac{\sum (\Delta m_{J,i}-\Delta m_J)(\Delta m_{H,i}-\Delta m_H)}
     {\sqrt{\sum (\Delta m_{J,i}-\Delta m_J)^2  
            \sum (\Delta m_{H,i}-\Delta m_H)^2} } \;\;\;,  \label{eq_r}
\end{equation}
where $\Delta m_\lambda$ ($\lambda=J$ or $H$) is the unweighted average 
of variability data in each band.  Table 4 tabulates the values of 
$r_{JH}=+0.74$, $r_{HK'}=+0.81$, and $r_{JK'}=+0.71$ for the entire 
sample, for which the 68.3\% confident interval is about 0.1 or less.  
The coefficient is higher than +0.59 for the radio-quiet sample, and even 
higher than +0.8 for the radio-loud sample.

Figure 8 shows the $z$-dependence of $r_{JH}$, $r_{HK'}$, and $r_{JK'}$.  
For the radio-quiet sample, all these coefficients at $z<0.1$ are equally 
high.  While $r_{JH}$ keeps a high value irrespective of $z$, the
coefficients $r_{JK'}$ and $r_{HK'}$ are getting smaller for higher $z$.  
This trend may have occurred from an  
underestimation of the correlation, because of lower statistical 
accuracy in the high-$z$ sample of smaller size, and because of 
larger errors in $\Delta m$ for the high-$z$ sample consisting mainly 
of faint AGNs. For the radio-loud sample, all the 
coefficients keep a high value irrespective of $z$ and their accuracy 
becomes higher for higher $z$, in sharp contrast with the radio-quiet 
sample.

The equivalence of $r_{JH}$, $r_{HK'}$, and $r_{JK'}$ was tested 
against the estimated true value of coefficient $r_{\rm true}$.   
Table 5 tabulates the estimation of $r_{\rm true}$ together with 
$P$ which represents the reliability of 
rejecting the hypothesis that $r_{JH}$, $r_{HK'}$, and $r_{JK'}$ are 
equivalent to each other.  For the radio-quiet sample, their 
equivalence at $z=0.1-0.3$ is significantly rejected, while not  
definitely so at higher $z$.  For the radio-loud sample, the equivalence 
at $z>0.3$ is rejected with negligible reliability, in other words, 
it is very probable that $r_{JH}$, $r_{HK'}$, and $r_{JK'}$ are the 
same at $z>0.3$. 
 
It is important to note that  $r_{JH}$, $r_{HK'}$, and $r_{JK'}$ for 
the radio-quiet sample are equally high at $z<0.1$, and $r_{HK'}$ is 
higher than $r_{JH}$ and $r_{JK'}$ at $z=0.1-0.3$.  In 
dust reverberation model, the NIR flux of AGNs is mainly emitted from 
hot dust which is heated up to the evaporation temperature $T_{\rm 
evap}\approx 1500$K (e.g., Kobayashi et al. 1993).  The equivalence 
of $r_{JH}$, $r_{HK'}$, and $r_{JK'}$ at $z<0.1$ can be explained by 
the black body radiation with constant $T_{\rm evap}\approx 1500$K  
dominating on the longer-wavelength region of the 1$\micron$ minimum 
seen in the rest-frame SED of AGNs and quasars.  On the other hand, 
the main variable component dominating on the shorter-wavelength 
region of the 1$\micron$ minimum is considered to be a power-law 
component, and its variability is not necessarily synchronized with 
the NIR variability.  Figure  9 shows this non-synchronization between 
UV/optical and NIR variabilities on either sides of 1$\micron$ in rest frame.
Thereby, at $z=0.1-0.3$, the correlation of variabilities between the 
$J$ band and longer wavelengths becomes weak, because the 1$\micron$ 
minimum in the rest frame moves to the $J$ band and the power-law 
variable component affects the flux there.  At $z>0.3$, the 1$\mu$m 
minimum moves in between the $H$ and $K'$ band.  In such case, 
$r_{JH}$ is kept high, determined mainly by the the power-law 
variable component, while $r_{HK'}$ and $r_{JK'}$ become low, 
determined by both the power-law and black-body components.

For the radio-loud sample, however, the obseved trends of $r_{JH}$, 
$r_{HK'}$, and $r_{JK'}$ are very different from those for the 
radio-quiet sample and may be explained by a mechanism of variability
other than dust reverberation. 

\subsection{The time scale of AGN variability}\label{sub_time_scale}

In this section we consider the relation between the time scale 
and other parameters that characterize the variability of AGNs.  
For pedagogical purpose, we introduce two representative 
functions to be fitted to our NIR data:
\begin{equation}
   \Delta m(A,p)=A(\Delta t_{\rm rest})^p \;\;\;,  
\label{eq_timescale_V1}
\end{equation}
and
\begin{equation}
   \Delta m(B,\tau)=B(1- \exp (-\Delta t_{\rm rest}/\tau)) \;\;\;.
\label{eq_timescale_V2}
\end{equation}
These functions correspond to a divergent increase of variability 
and an asymptotic increase with increasing $\Delta t_{rest}$,
respectively.

Tables 6a and 6b tabulate the fitted values of the parameters 
and their 95\% and 99\% confidence intervals, applied to the AGNs 
in the entire sample as well as the radio-quiet and radio-loud samples.
The curves with such  fitted values are shown in Fig. 10.
We note that $p$ and $\tau$ are searched for in their postive values 
to avoid division by 0 at $\Delta t_{\rm rest}=0$ in equations 4 
and 5. If their fitted values or the lower limits of their confidence 
intervals became negative, they were given the value of zero.

All the fitted curves are accepted independent of how the data are 
divided into subgroups.  Thus, it is difficult to prefer the 
function in equation 4 to the other in equation 5.  However, it is 
generally suggested from Tables 6a and 6b that $p$ and $\tau$ for 
the radio-loud sample are smaller 
than those for the radio-quiet sample, which indicates that the  time 
scale of NIR variability of radio-loud AGNs  
is shorter than that for the radio-quiet AGNs.
  
Figure 11 shows the fitted values of  $p$ 
and $\tau$ in comparison with those obtained by Cristiani et al. (1996) 
from the optical variability data in the literature.  The vertical
errorbar is for the 68.3\% confidence interval, and the horizontal 
errorbar for the covered range of wavelengths. It is interesting to 
see that the time scale of optical variability by Cristiani et al. 
(1996) agrees, within the  range of uncertainties,  with the time scale 
of NIR variability for the radio-quiet sample.  On the contrary, this  
is not the case for the radio-loud sample, where  the time scale 
of optical variability is significantly longer than the time scale 
of NIR variability.

From the viewpoint of dust reverberation, it is natural that 
the NIR variability occurs on a rather elongated time scale, because 
the UV/optical variability 
is re-emitted in the NIR from a 
more extended region of dust 
producing a spread of arrival times.  
Consequently, whether the NIR variability occurs as fast as or 
faster than the UV/optical variability gives a clue which limits or 
rejects  a mechanism of variability based  upon  the dust
reverberation model. 
In this regard, it is difficult to explain by the dust reverberation 
model how the NIR variability of the radio-loud AGNs occurs faster 
than the optical variability, as seen from the comparison in Fig. 11.  

\subsection{The relation between  near infrared variarility 
        and radio activity}\label{sub_kikou}

We have highlighted the difference of NIR variability between the
radio-quiet and radiou-loud AGNs.  Obtained features for these 
respective samples are summarized as follows:

\noindent
(1) The amplitude of ensemble NIR variability $\Delta m$ in the 
radio-quiet sample is smaller than in the radio-loud sample.

\noindent
(2) The wavelength-dependence of $\Delta m$ in the NIR region 
is not found in both the radio-quiet and radio-loud samples.

\noindent
(3) No correlation is found between $\Delta m$ and $M_B$ in 
the radio-quiet sample, but a negative correlation is suggested 
if corrected for the effect of possible contamination of a host 
galaxy component.  On the other hand, a positive correlation is 
found between $\Delta m$ and $M_B$ in the radio-loud sample.

\noindent
(4) The correlation coeficient $r_{HK}$ between the $H$ and 
$K$ variabilities at $z=0.1-0.3$ is significantly higher than 
$r_{JH}$ and $r_{JK'}$ in the radio-quiet sample.  On the 
other hand, the coeficients $r_{JH}$, $r_{JK'}$ and $r_{HK'}$ 
at $z>0.3$ have a high value of $0.9-0.95$ in the radio-loud 
sample.

\noindent
(5) The time scale of NIR variability in the radio-quiet sample 
is longer than in the radio-loud sample. 

The features of 2, 3, and 4 for the radio-quiet AGNs are 
consistent with those expected from a mechanism of variability 
by  dust reverberation.  Furthermore, the feature of 5 for the 
radio-quiet AGNs, if combined with the time scale of 
optical variability by Cristiani et al. (1996), is also 
consistent with dust reverberation.
 
However, the features of 2 to 5 for the radio-loud AGNs are not 
explained consistently by  dust reverberation.  A non-thermal 
variable component, as a substitute of thermal radiation from 
hot dust, is worth considering. For example, Sanders et al. 
(1989) proposed such a non-thermal component which would vary 
more strongly and faster in the compact region than the thermal 
radiation  from an extended hot dust torus.

We  proceed with the working hypothesis such that dust 
reverberation is responsible for the emission and variability of 
radio-quiet AGNs, while a non-thermal variable component is 
responsible for the emission and variability of radio-loud AGNs.  
In the next subsections, we check the plausibility of this 
hypothesis using additional data from the literature.

\subsubsection{The radio strength of the dust reverberation sample}

It is known that the time  delay of NIR variability lagged behind the 
the UV/optical variability is a key prediction of the 
dust reverberation model.  Therefore, measurements of UV/optical and 
NIR light curves, based on multi-wavelength monitoring observations 
of AGNs, immediately reject or justify the application of the dust 
reverberation model to individual AGNs. 

The AGNs with measured time delay, to which the dust reverberation
model  
is successfully applied, are taken from the literature, and their 
data of $z$, $M_B$, and 
$f_{\nu}({\rm 6cm})/f_{\nu}(V)$ are summarized in Table 7.  
It is evident that these AGNs are 
classified as radio-quiet with $f_{\nu}({\rm 6cm})/f_{\nu}(V)<10$. 
The sample of these AGNs is heterogeneous and is far from complete. 
It is possible that the sample is biased in favor of nearby and/or 
radio-quiet AGNs.

The first systematic monitoring observations of many Seyfert 1 AGNs 
and quasars in the $V$ and $K$ bands were performed by Nelson 
(1996a).  Table 8 shows the statistics based 
on Nelson's sample consisting of 51 program AGNs.  The first row 
represents the numbers of radio-quiet and radio-loud AGNs in his 
entire sample of 51 AGNs, the second row for the subsample of 33 
AGNs which were found to vary in both the $V$ and $K$ bands, and 
the third row for the subsample of 6 AGNs for which the time delay 
of the $K$-band variability relative to the $V$-band variability 
were measured. 

We see  that the AGNs with measured time delay, 
which are best explained by the simple dust reverberation model, 
are always classified as radio-quiet.  In other words, radio-quiet 
AGNs are potential targets for multi-wavelength monitoring from 
which the time delay between the NIR and UV/optical variabilities 
can certainly be measured. 

\subsubsection{The relation between variability and flatness of radio SED}

The large and fast NIR variability of radio-loud AGNs found in this 
paper is not explained by dust reverberation model.  In order to 
understand what causes such variability, we further classify  the 
radio-loud AGNs with respect to their spectral feature in the radio region,
being either flat ($\alpha>-0.5$) or steep ($\alpha<-0.5$), if fitted to a 
power-law 
form of $f_\nu \propto \nu^\alpha$ (Sanders et al. 1989).  

The values of power index $\alpha$ in our sample were determined using 
the data of radio fluxes at  6 cm and 11 cm taken from the VV catalog.
Figure 12 shows that $\alpha$ is not correlated with $M_B$ or $z$.   
Figure 13 shows that the ensemble variability 
in the $J$, $H$, and $K'$ bands is $\Delta m\approx 0.35-0.45$ for the 
radio-loud and flat-spectrum  AGNs, which is systematically 
higher than 
$\Delta m\approx 0.2$ for the radio-loud and steep-spectrum AGNs.  
It is therefore the flat component that actually brings about the large 
and fast NIR 
variability. 

The radio spectral index, $\alpha$, is known 
to be well correlated with the radio-loud AGN 
morphology;  while the steep spectrum is associated with extended 
lobe-dominant 
sources, the flat spectrum is associated with core-dominant variable sources 
such as OVVs, highly polarized QSOs, and BL lacs.   Thereby, it is 
reasonable to
conclude that a non-thermal variable component, as exemplified by non-thermal 
emissions from such objects (e.g., Robson et al. 1993; Bloom et al. 1994),
is responsible for occurrence of  features of the NIR variability found for 
the radio-loud AGNs in this paper. 

\section{Summary}

We presented  comprehensive study of NIR variability of 226 AGNs based on
multiple observations in the $J$, $H$, and $K'$ bands.   Our sample 
consists mainly of Seyfert 1 AGNs and QSOs. About a quarter of objects 
in each category are radio loud.  
The AGNs in the entire sample have redshifts spanning  a range 
from $z=0$ to 1, and the absolute $B$-magnitudes from $M_B=-29$ to $-18$.  

Based on the method of differential photometry, 
we find that the ensemble NIR variability 
for the entire sample of AGNs is typically $\Delta m\approx 0.2$ mag.  
When the sample is divided by radio strength, 
the variability for the radio-quiet sample is systematically
smaller than that for the radio-loud sample.  No clear 
wavelength-dependence of 
$\Delta m$ in the 
NIR region is found for either the radio-quiet or radio-loud sample, in 
sharp contrast with the 
UV/optical result in the literature.  

We examined the dependence of $\Delta m$ on various quantities such as 
radio strength, $M_B$, and $z$, with special attention 
as to whether their ensemble variability would 
support or reject the simple-minded dust reverberation model for AGNs. 

The radio-quiet AGNs show no significant correlation between $\Delta m$ 
and $M_B$, although negative correlation is suggested if corrected for 
the effect of possible contamination by a host galaxy component.  
On the other hand, the radio-loud AGNs show a positive correlation 
between $\Delta m$ and $M_B$.

The radio-quiet AGNs give a significantly higher correlation coefficient 
$r_{HK}$ between the $H$ and $K$ variabilities at $z=0.1-0.3$, when 
comapred to  $r_{JH}$ and $r_{JK'}$.   On the other hand, the radio-loud
AGNs give  a high value of $0.9-0.95$ to all the coefficients $r_{JH}$, 
$r_{JK'}$ and $r_{HK'}$ at $z>0.3$.

Time development of ensemble variability is examined using heuristic
functions.  The time scale of NIR variability for the radio-quiet AGNs in 
this paper 
is not shorter than the time scale of UV/optical variability given in the 
literature.  
However, the time scale of NIR variability of the radio-loud AGNs 
is significantly shorter than the time scale of their UV/optical variability.

All the features of NIR variability for the  radio-quiet AGNs are 
consistent with those expected from the dust reverberation model.  
However, it is difficult for this simple-minded model to explain 
the features of  the radio-loud AGNs, and a non-thermal variable component 
is suggested as a viable candidate for  causing the large and fast NIR 
variability of the radio-loud AGNs.


We are grateful to H. Okuda, M. Narita and other staff of the infrared 
astronomy group of  the Institute of Space and Astronautical Science 
(ISAS) for their support in using their 1.3m telescope. We thank the 
staff of the Advanced Technology Center of the National Astronomical 
Observatory of Japan (NAOJ) for their new coating of the mirror 
of the 1.3m telescope at the ISAS.  Gratitude is also extended to the 
Computer Data Analysis Center of the NAOJ.  This work has made use of 
the NASA/IPAC Extra Galactic Database (NED), and has been supported 
partly by the Grand-in-Aid (07CE2002, 10304014) of the Ministry of 
Education, Science, Culture, and Sports of Japan and by the Torey 
Science Foundation.

\appendix

\section{estimation of the ensemble variability}\label{apnd_V}

This section describes how we  estimate  variability and its error 
in this work using the data for which the sample number and accuracy 
are both limited.
The data obtained in this work are given in the 
format of $(\Delta m_1, \sigma_{1})$, 
$(\Delta m_2, \sigma_{2})$ ... 
$(\Delta m_N, \sigma_{N})$
for the sample of AGN$_1$, AGN$_2$ ... AGN$_N$.
Here, $\Delta m_i$ and $\sigma_{i}$ represent the variability and its 
error of the $i$-th object.  Hereafter, the real variability of the 
$i$-th object is described as $v_i$, and the difference between $v_i$ 
and  $\Delta m_i$ is given by $e_i$.  Then, $e_i$ can be regarded as 
a random variable described by the normal distribution having zero
average and the standard deviation $\sigma_i$, that is,
\begin{eqnarray}
\Delta m_i=v_i+e_i , \hspace{5mm} <e_i>=0 , 
\hspace{5mm} <e^2_i>=\sigma_i ,  \label{eq_e1e2}
\end{eqnarray}
where $<>$ represents the expectation.  The real variability 
of the $i$-th object $v_i$ is assumed to be a random variable that follows 
the normal distribution with zero average and the standard deviation 
$v_0$ independent of $i$.

In the ideal case where the data have zero error and 
the sample size is infinite, 
the dispersion of variability is given by 
\begin{eqnarray}
  \lim_{N \rightarrow \infty} \left[\frac{ \sum^N_{i}v^2_i}{N} \right] 
\label{eq_ideal}
\end{eqnarray}
Hereafter, the aim is to estimate the value of equation \ref{eq_ideal} 
and the error from the existing data. 
Consider 
\begin{eqnarray}
 v^2_i=\Delta m^2_i-2v_ie_i-e^2_i.
\end{eqnarray}
Then, the expectation of the second term in the right-hand side vanishes
because $v_i$ and $e_i$ are independent of each other.  Therefore, the 
expectation of $v^2_i$ becomes 
\begin{eqnarray}
  <v^2_i>=\Delta m^2_i-\sigma^2_i \label{eq_v2}.
\end{eqnarray}
{\noindent}
Substitution of the observed data into equation \ref{eq_ideal} gives 
\begin{eqnarray}
   \lim_{N \rightarrow \infty} \left[\frac{ \sum^N_{i}v^2_i}{N} \right]
  \sim \frac{ \sum^N_{i}v^2_i}{N} 
  \sim \frac{\sum^N_i \Delta m^2_i - \sum^N_i \sigma^2_i}{N}\label{eq_ap_sf2}
   \equiv \overline{v^2_0},
\end{eqnarray}
where the first $\sim$ stems from the finite sample number and the 
second $\sim$ stems from the error of the data.  The overline of 
$\overline{v^2_0}$ indicates that the value is the expectation based on
the exisiting data.

The error of equation \ref{eq_ap_sf2} is estimated as follows:  First, we 
focus on the error caused by the deviation of the ideal expectation 
from that of the actual data with finite sample number.  It should be 
noticed that $\sum^N_i(v_i/v_0)^2$ is the random variable described 
by the $\chi^2$ distribution with $N$ degrees of freedom.  Since 
$v_i/v_0$ $(i=1,2...N)$ is a  random variable that follows the 
normal distribution with  zero average and the standard deviation 
of unity, the expectation $\sum^N_i(v_i/v_0)^2$ is equal to $N$ with
the dispersion of $2N$.  Therefore, the error of equation \ref{eq_ap_sf2} 
arising from the finite sample number is estimated as 
$\pm \sqrt{2v^4_0/N}$.

We next focus on the error caused by the deviation of the ideal  
expectation from that of the actual data with non-zero error.
Assuming that $v_i$ and $e_i$ are independent of each other, the 
difference between $v^2_i$ and $<v^2_i>$ is calculated as
\begin{eqnarray}
<(v^2_i-<v^2_i>)^2>&=&<(\Delta m^2_i-2e_iv_i-e^2_i-\Delta 
m^2_i+\sigma^2_i)^2> \nonumber \\
&=&<4e^2_iv^2_i+e^4_i+\sigma^4_i+4e^3_iv_i-4e_iv_i\sigma^2_i-2e^2_i\sigma^2_
i>  \nonumber \\
&=& 4\sigma^2_i<v^2_i>+<e^4_i>-\sigma^4_i   \label{eq_disp}, 
\end{eqnarray}
where equation \ref{eq_e1e2} is used.  Using equation \ref{eq_v2}
and $<e^4_i>=3\sigma^4_i$, equation \ref{eq_disp} furthermore reduces to
\begin{eqnarray}
 <(v^2_i-<v^2_i>)^2>&=& 4\sigma^2_i<v^2_i>+2\sigma^4_i \nonumber \\
                    &=& 4\sigma^2_i \Delta m^2_i - 2\sigma^4_i .
\end{eqnarray}
Therefore, the error of equation \ref{eq_ap_sf2} arising from the non-zero 
error of the variability data is estimated as 
$\pm \sqrt{\sum^N_i(4\Delta m^2_i\sigma^2_i-2\sigma^4_i)/N^2}$. 
Using equation \ref{eq_ap_sf2} and the above error estimates, the expectation
value of $v^2_0$ and its error are estimated as
\begin{eqnarray}
\overline{v^2_0}
&=&  \frac{\sum^N_i \Delta m^2_i - \sum^N_i \sigma^2_i}{N}  
     \pm  \sqrt{\frac{2v^4_0}{N}+\frac{\sum^N_i(4\Delta 
m^2_i\sigma^2_i-2\sigma^4_i)}{N^2} } \nonumber \\
&\sim& \frac{\sum^N_i \Delta m^2_i - \sum^N_i \sigma^2_i}{N}
      \pm  \sqrt{\frac{2(\sum^N_i \Delta 
m^2_i-\sum^2_i\sigma^2_i)^2}{N^3}+\frac{\sum^N_i(4\Delta 
m^2_i\sigma^2_i-2\sigma^4_i)}{N^2} }  .
\end{eqnarray}
Finally, using the usual error propagation, the expectation value of $v_0$ 
and its error are estimated as
\begin{eqnarray}
\overline{v_0} &=&   \sqrt{\frac{\sum^N_i \Delta m^2_i - \sum^N_i 
\sigma^2_i}{N}} \nonumber \\
  &\pm& \frac{1}{2} \left[\frac{\sum^N_i \Delta m^2_i - \sum^N_i 
\sigma^2_i}{N} \right]^{-1/2} 
       \sqrt{\frac{2(\sum^N_i \Delta 
m^2_i-\sum^2_i\sigma^2_i)^2}{N^3}+\frac{\sum^N_i(4\Delta 
m^2_i\sigma^2_i-2\sigma^4_i)}{N^2} } .
\end{eqnarray}



\clearpage

\figcaption{The ensemble variability of AGNs in the $J$, $H$ and $K'$ bands.
        Shown are the results for all AGNs, and radio-quiet and loud
        AGNs ({\it upper panel}),  for radio-quiet AGNs divided by 
        $\Delta t_{\rm rest}$ ({\it lower left panel}),
        and for radio-loud AGNs 
        divided by $\Delta t_{\rm rest}$ ({\it lower right panel}). 
        Only the data with more than two reference objects
        and with accuracy higher than 0.1mag are plotted.
}

\figcaption{The ensemble variability of radio-quiet AGNs in the $J$, $H$ and 
        $K'$ bands.  Shown are the results for short-$\Delta t_{\rm rest}$
             AGNs divided by $M_B$ ({\it upper left panel}) or by $z$
             ({\it lower left panel}),  and  those for long-$\Delta 
t_{\rm rest}$
             AGNs divided by $M_B$ ({\it upper right panel}) or by $z$
             ({\it lower right panel}).
}

\figcaption{The ensemble variability of radio-loud AGNs in the $J$, $H$ and 
        $K'$ bands.  Shown are the results for short-$\Delta t_{\rm rest}$
             AGNs divided by $M_B$ ({\it upper left panel}) or by $z$
             ({\it lower left panel}),  and  those for long-$\Delta 
          t_{\rm rest}$
             AGNs divided by $M_B$ ({\it upper right panel}) or by $z$
             ({\it lower right panel}).
}

\figcaption{The relation among characteristic parameters for radio-quiet AGNs.
            Shown are $\Delta m$, $\Delta t_{\rm rest}$, and $M_B$
            plotted against $z$
                    ({\it left column}), and  $\Delta m$, $\Delta t_{\rm 
            rest}$, and $z$ plotted 
                   against $M_B$ ({\it right column}). 
}

\figcaption{The relation among characteristic parameters for radio-loud AGNs.
     Shown are $\Delta m$, $\Delta t_{\rm rest}$, and $M_B$ plotted 
    against $z$ ({\it left column}), the  $\Delta m$, $\Delta t_{\rm 
    rest}$, and $z$ plotted   against $M_B$ ({\it right column}). 
}

\figcaption{The relation between variabilities in two NIR bands for 
radio-quiet AGNs divided by $\Delta t_{\rm rest}$.
}

\figcaption{The relation between variabilities in two NIR bands for 
radio-loud AGNs divided by $\Delta t_{\rm rest}$.
}

\figcaption{The coefficient of correlation $r_{ij}$  between AGN 
        variabilities in the $\lambda_i$ and $\lambda_j$ 
         bands as  a function of redshift $z$. Shown are the results 
        for radio-quiet AGNs
         ({\it upper panel}), and for radio-loud AGNs ({\it lower panel}).
}

\figcaption{The ensemble variability of AGNs as a function of
    rest-frame frequency. 
    Results in the NIR are from  this paper, and those in the 
    UV/optical are from the literature.   
    Filled small symbols are for 3 month-data
   ({\it triangles}, Cimatti et al. 1993;
   {\it filled circles}, Trevese et al. 1994; {\it squares},  
   De Clemente et al. 1996; {\it pentagons},  Hook et al. 1994; 
   {\it hexagons},  Cristiani et al. 1990). The open small symbols are
   the same as filled ones, but for 2yr-data. 
}

\figcaption{The ensemble variability of AGNs as a function 
       of rest-frame time interval between two observations.
       Shown are the results for all AGNs
        ({\it upper panel}) and for radio-quiet and loud AGNs 
({\it lower panel}). 
      The solid and dotted lines represent the best-fit  curves of 
      $\Delta m(A,p)=A(\Delta t_{\rm rest})^{p}$ and 
      $\Delta m(B,\tau)=B(1- \exp (-\Delta t_{\rm rest}/\tau))$,
            respectively.
}

\figcaption{The wavelength-dependence of fitted parameters to the ensemble 
        variability of AGNs, such as  the power index $p$ in 
            $\Delta m(A,p)=A(\Delta t_{\rm rest})^{p}$ ({\it Upper panel}),
            and the timescale $\tau$ in 
     $\Delta m(B,\tau)=B(1- \exp (-\Delta t_{\rm rest}/\tau))$ 
      ({\it Lower panel}).
       The vertical errorbars represent the 68.3\% confidence interval,
       and horizontal errorbars represent the range of wavelength coverage 
            in the analysis.  Results in the NIR are from  this paper, and 
            those in the optical are from Cristiani et al. (1996).
}

\figcaption{The distribution of radio-loud AGNs in the $\alpha$ versus $M_B$
    diagram ({\it upper panel}) and in the $\alpha$ versus $z$ diagram 
    ({\it lower panel}).  The boundary between the steep and flat spectra 
    is placed at $\alpha=-0.5$.  Open and filled circles correspond to 
    short- and long-$\Delta t_{\rm rest}$ samples, respectively.
}

\figcaption{The ensemble variability of radio-loud AGNs in the $J$, $H$ and 
       $K'$ bands.  Shown are the results for those with flat and steep 
       radio spectra.
}

\clearpage
\begin{figure}
  Figure 1\\
  \epsscale{0.5}
  \plotone{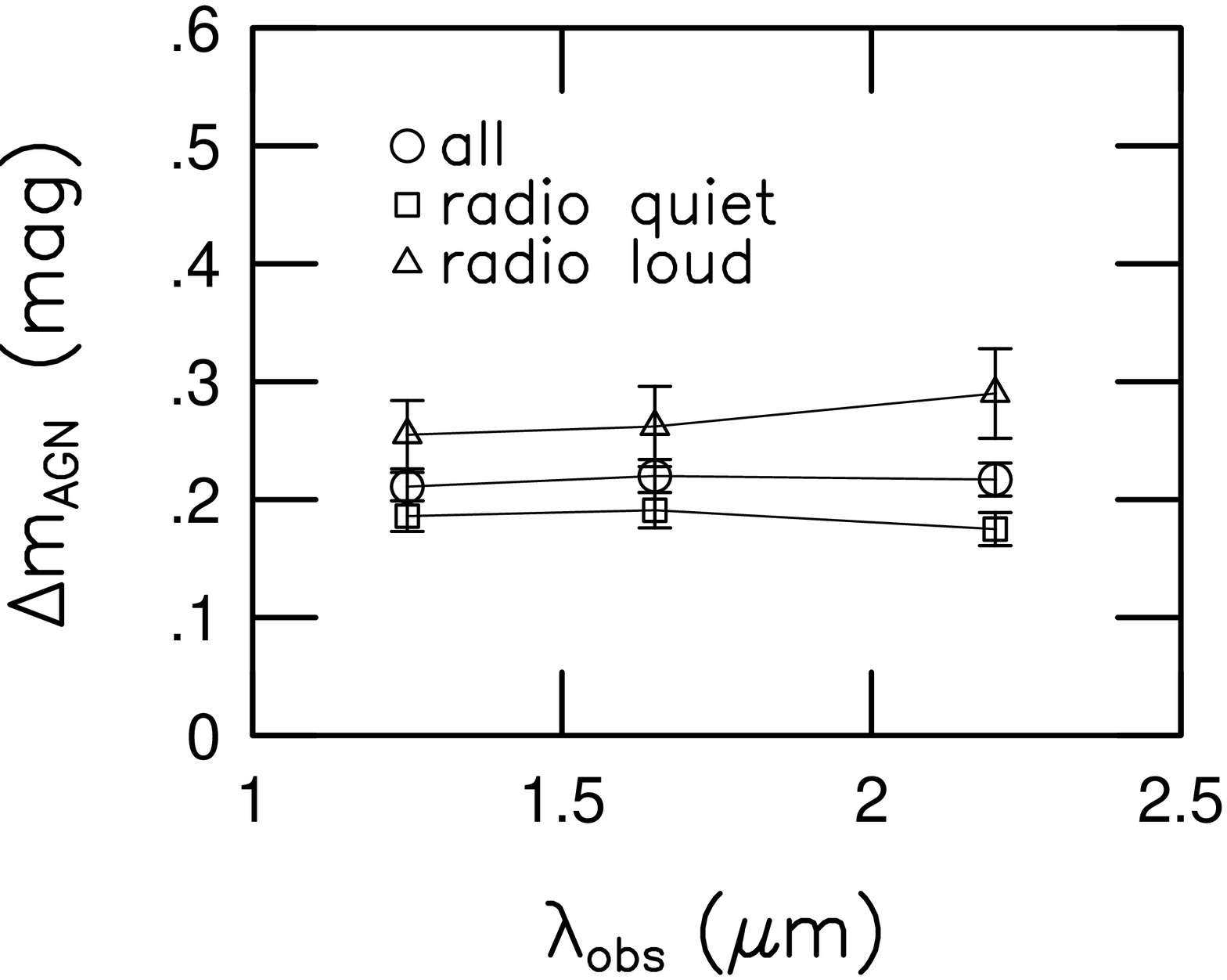}
\\
  \plotone{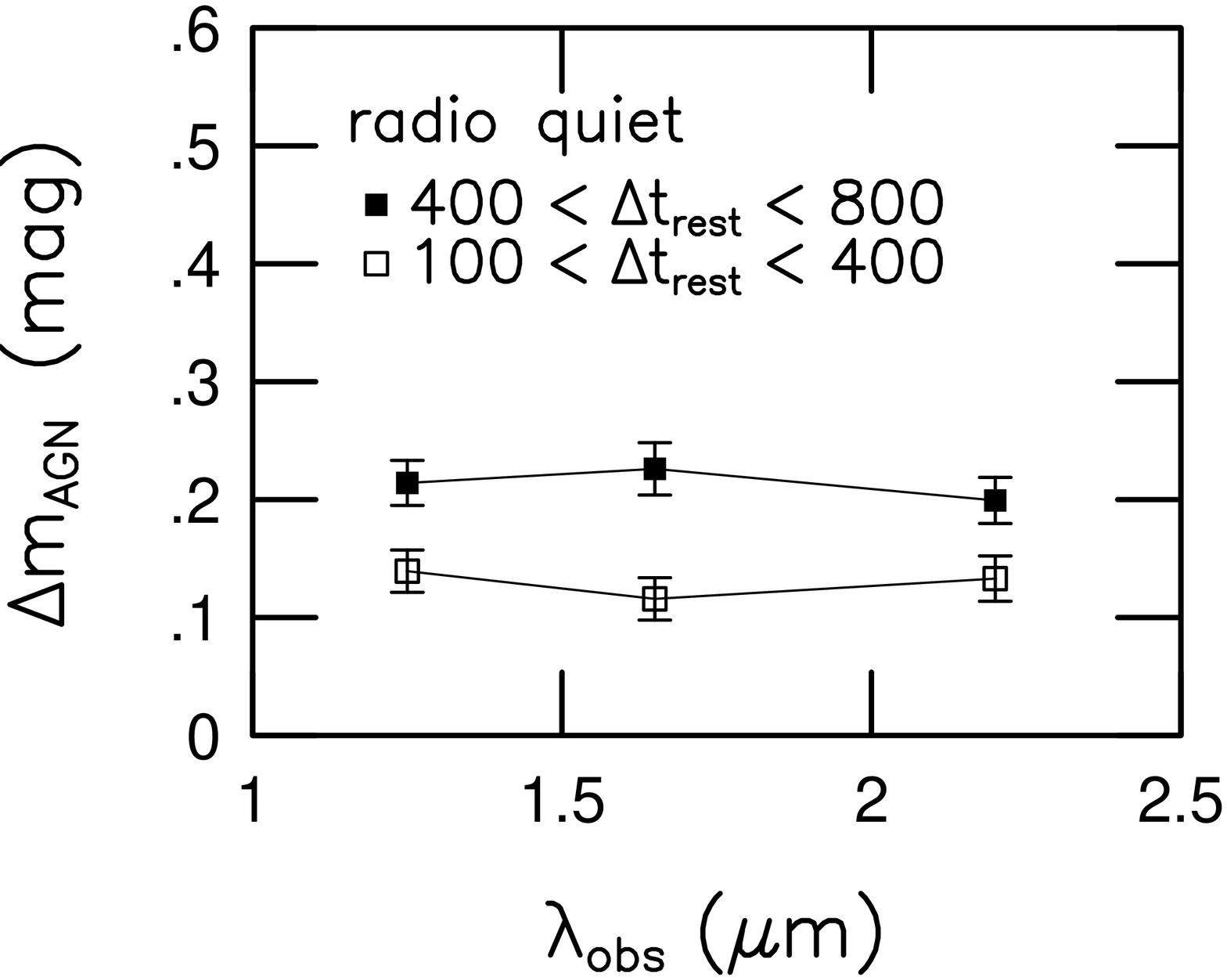}
  \plotone{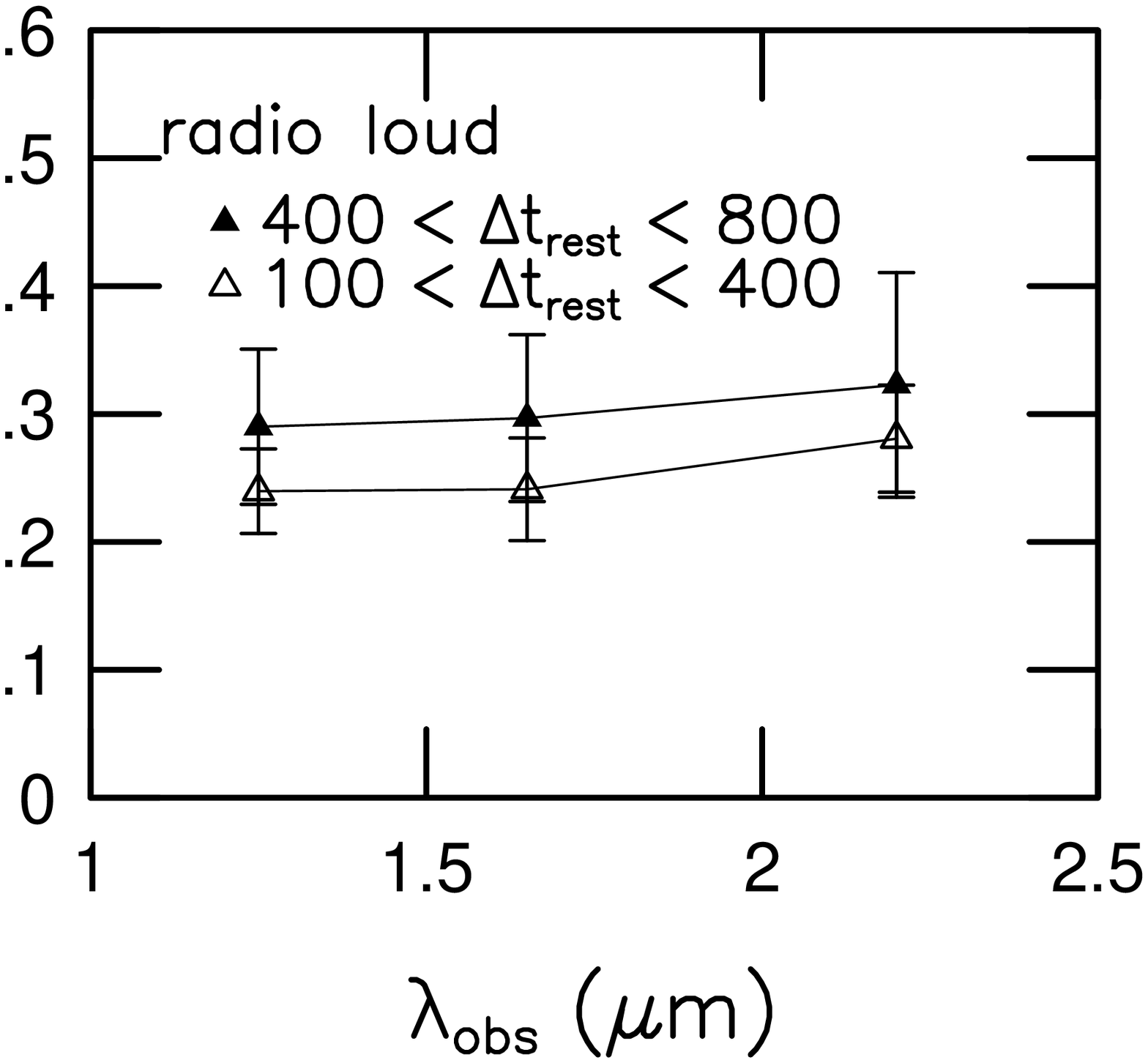}
\end{figure}

\clearpage
\begin{figure}
  Figure 2\\
  \epsscale{1}
  \plotone{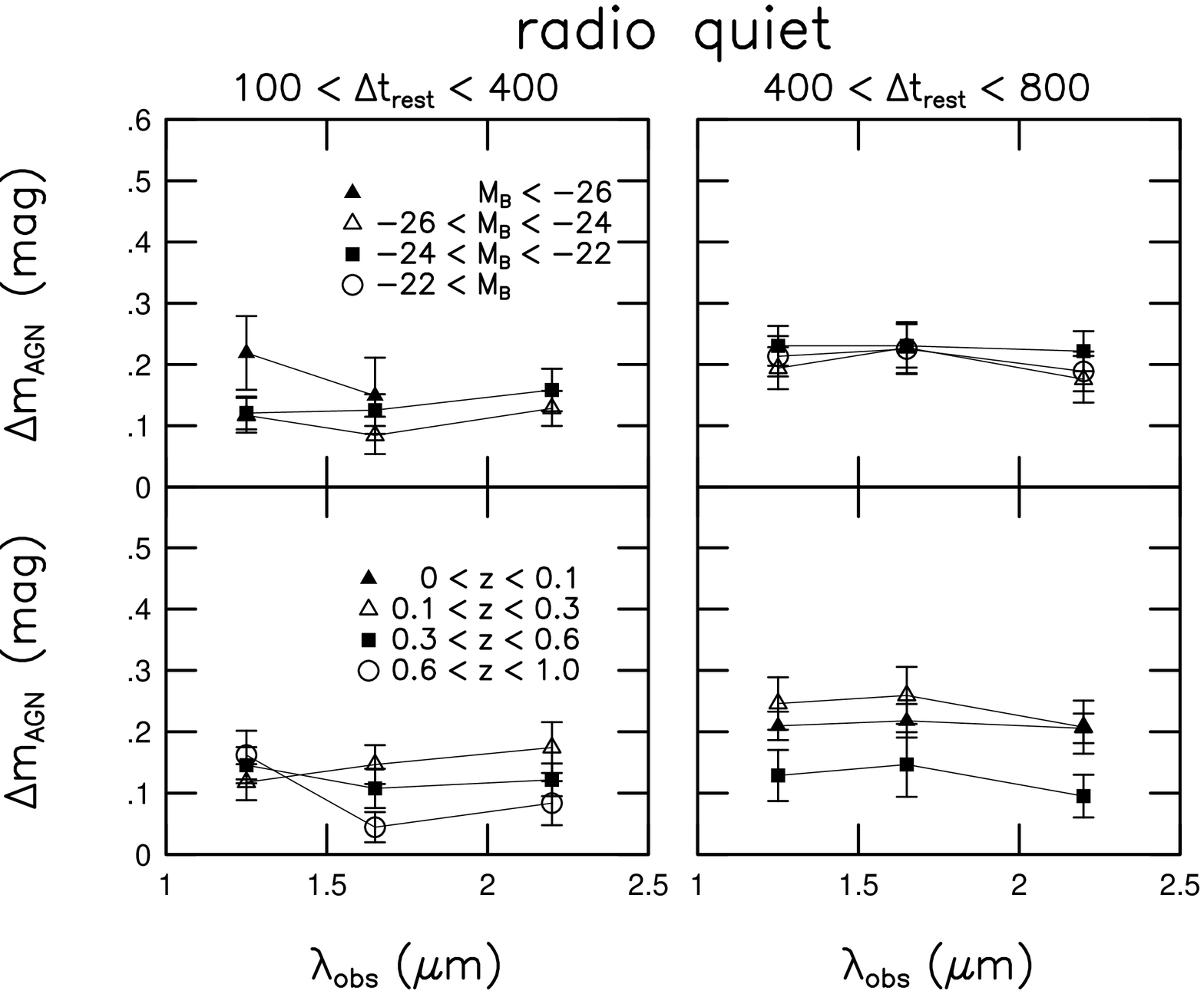}
\end{figure}

\clearpage
\begin{figure}
  Figure 3\\
  \epsscale{1}
  \plotone{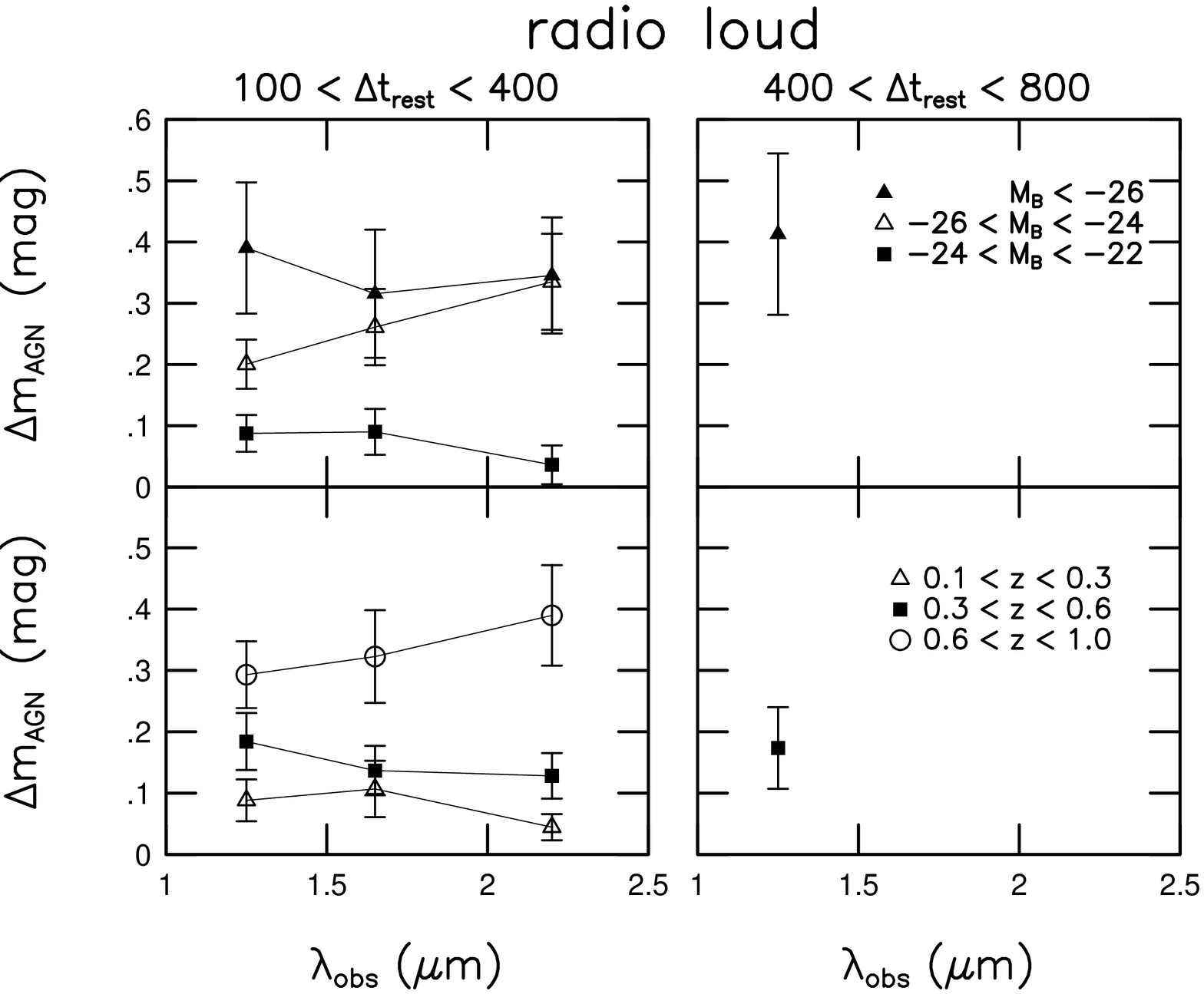}
\end{figure}

\clearpage
\begin{figure}
  Figure 4\\
  \epsscale{1}
  \plotone{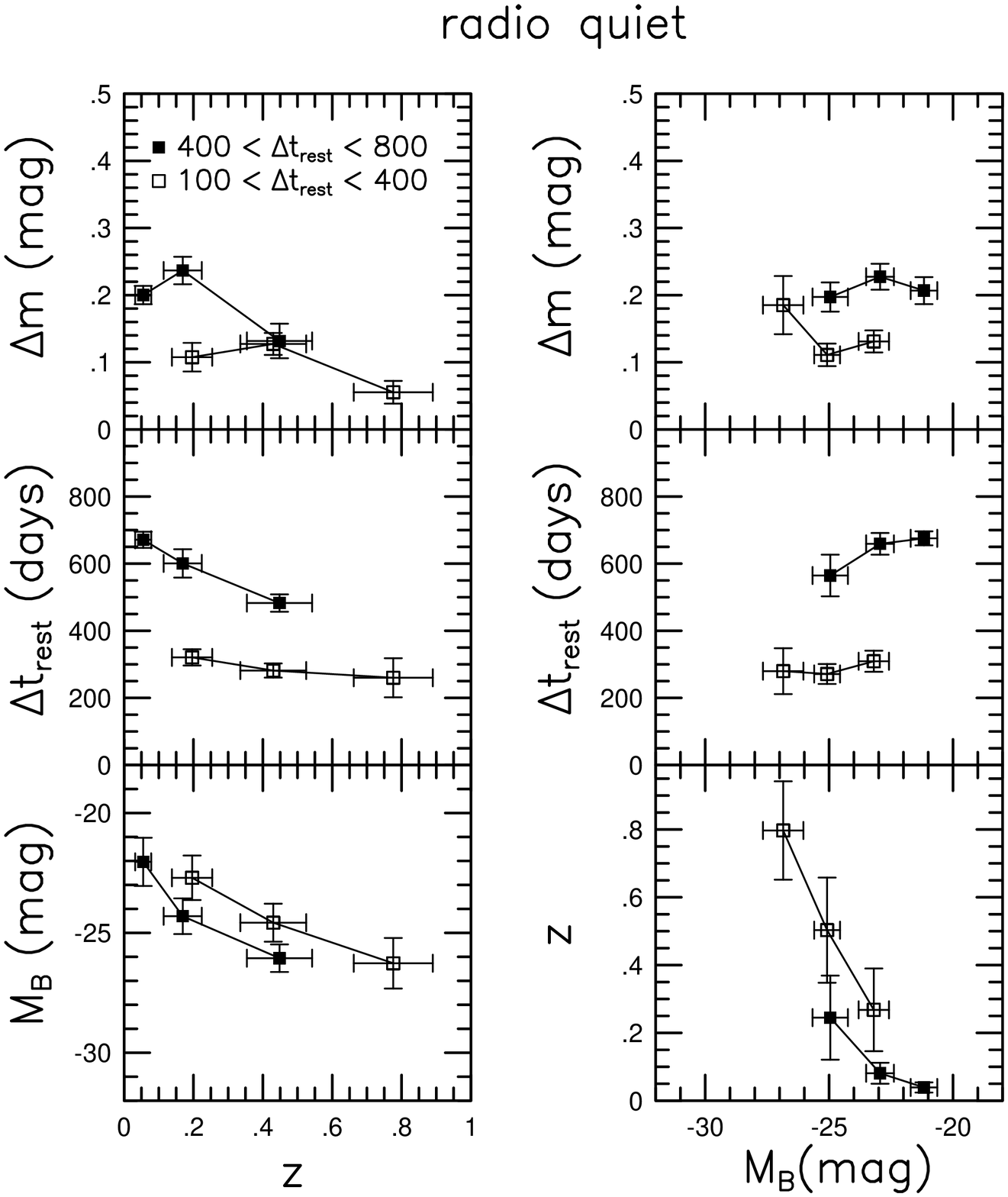}
\end{figure}

\clearpage
\begin{figure}
  Figure 5\\
  \epsscale{1}
  \plotone{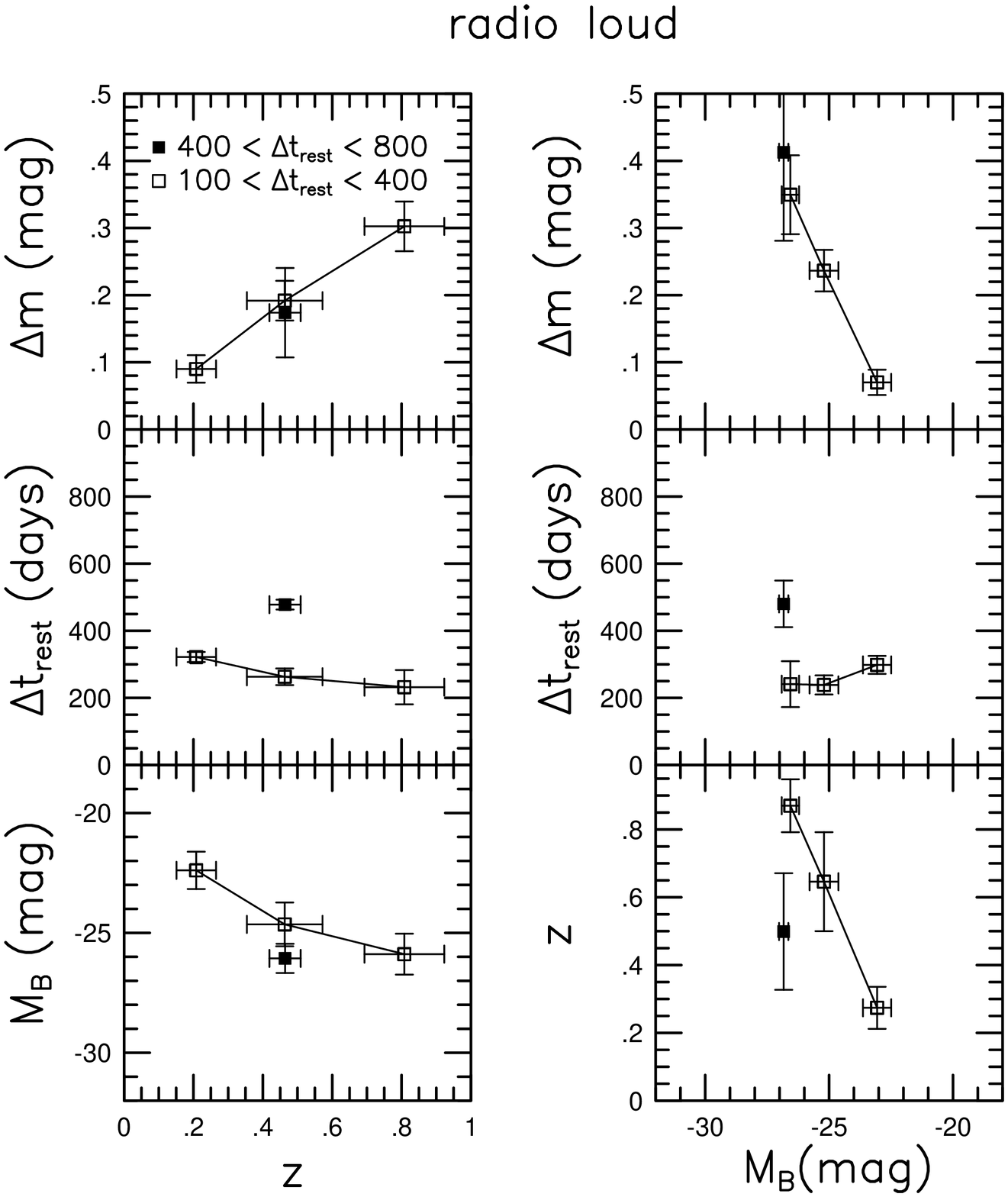}
\end{figure}

\clearpage
\begin{figure}
  Figure 6\\
  \epsscale{0.5}
  \plotone{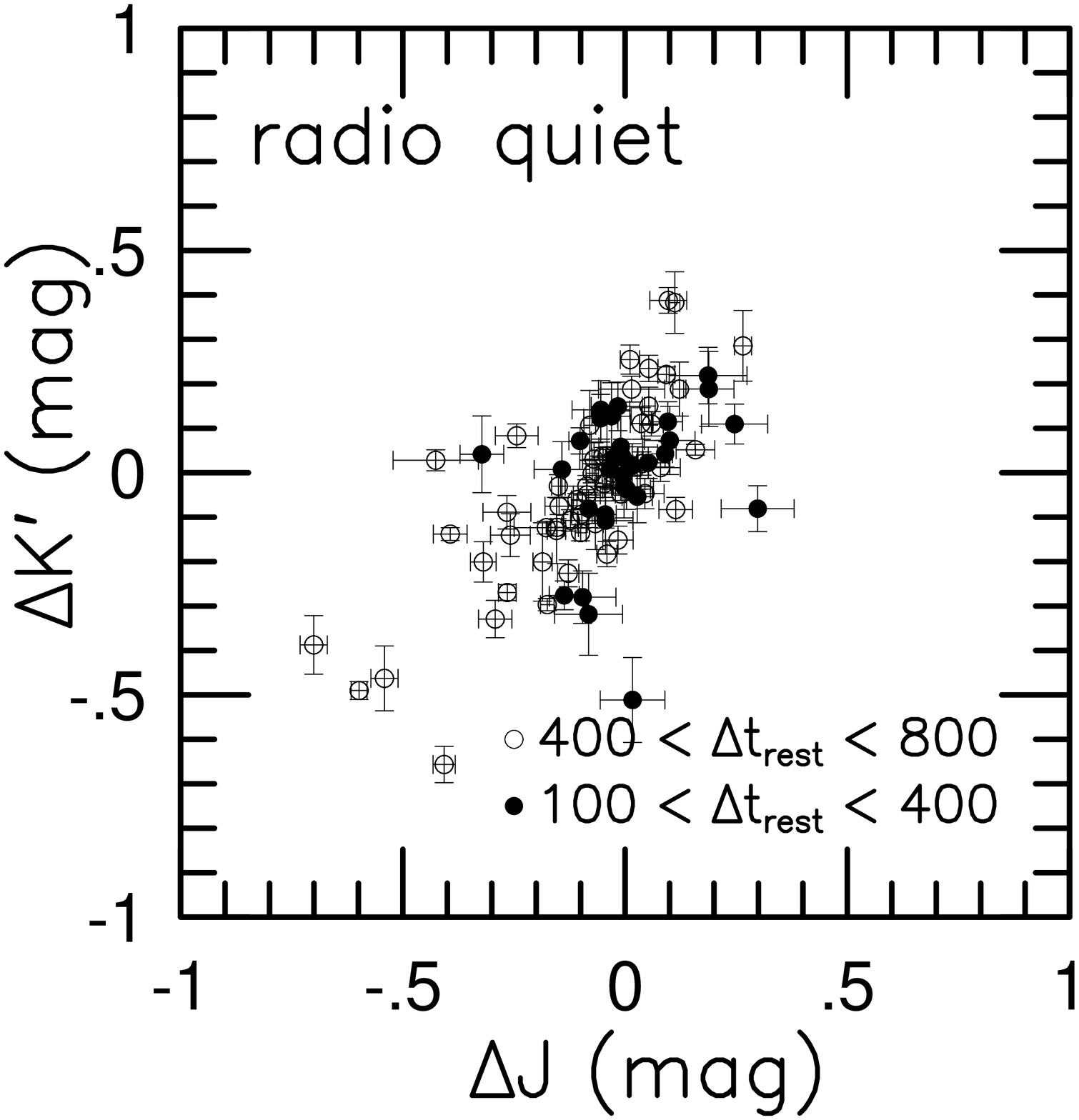}
\\
  \plotone{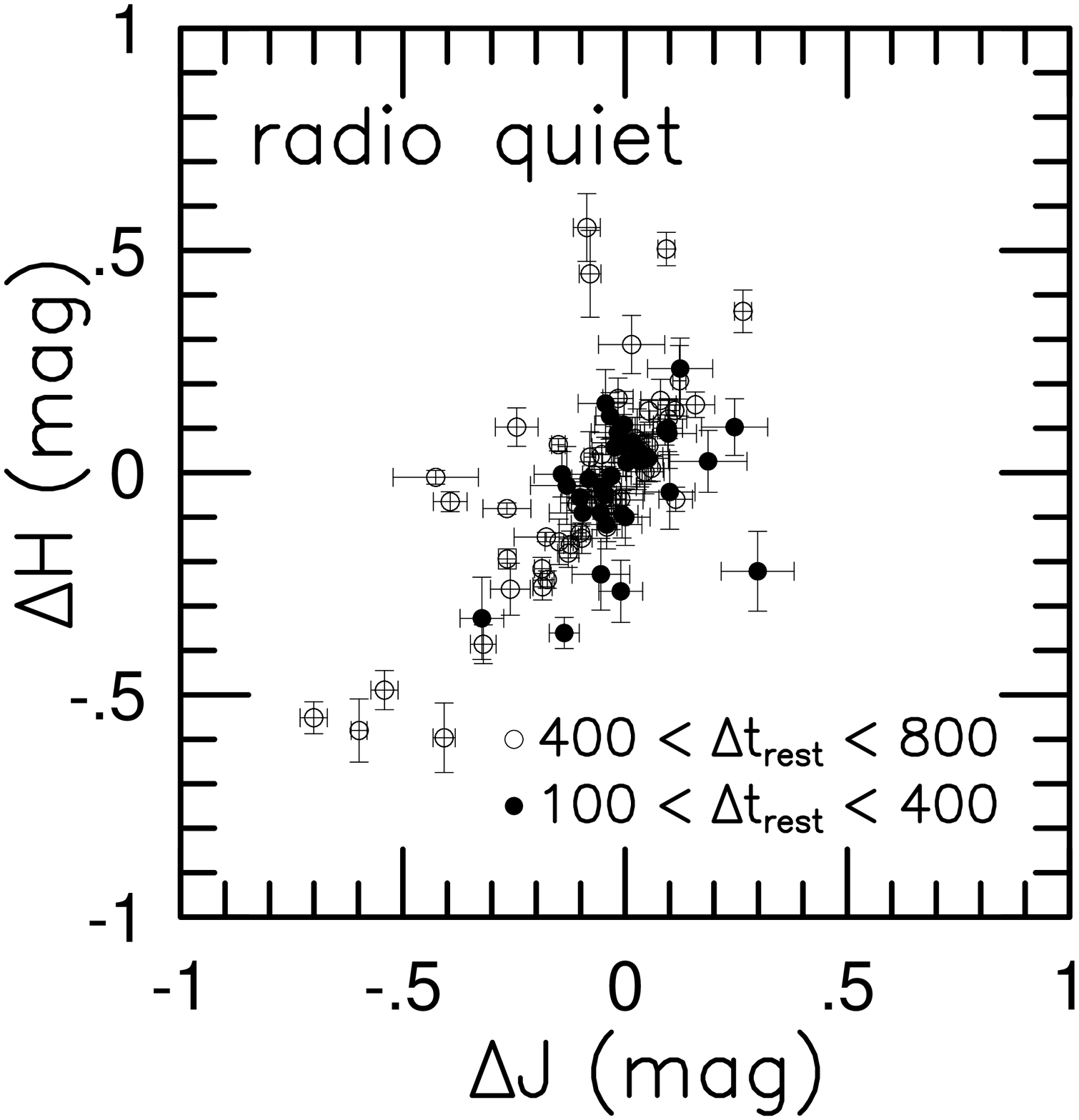}
  \plotone{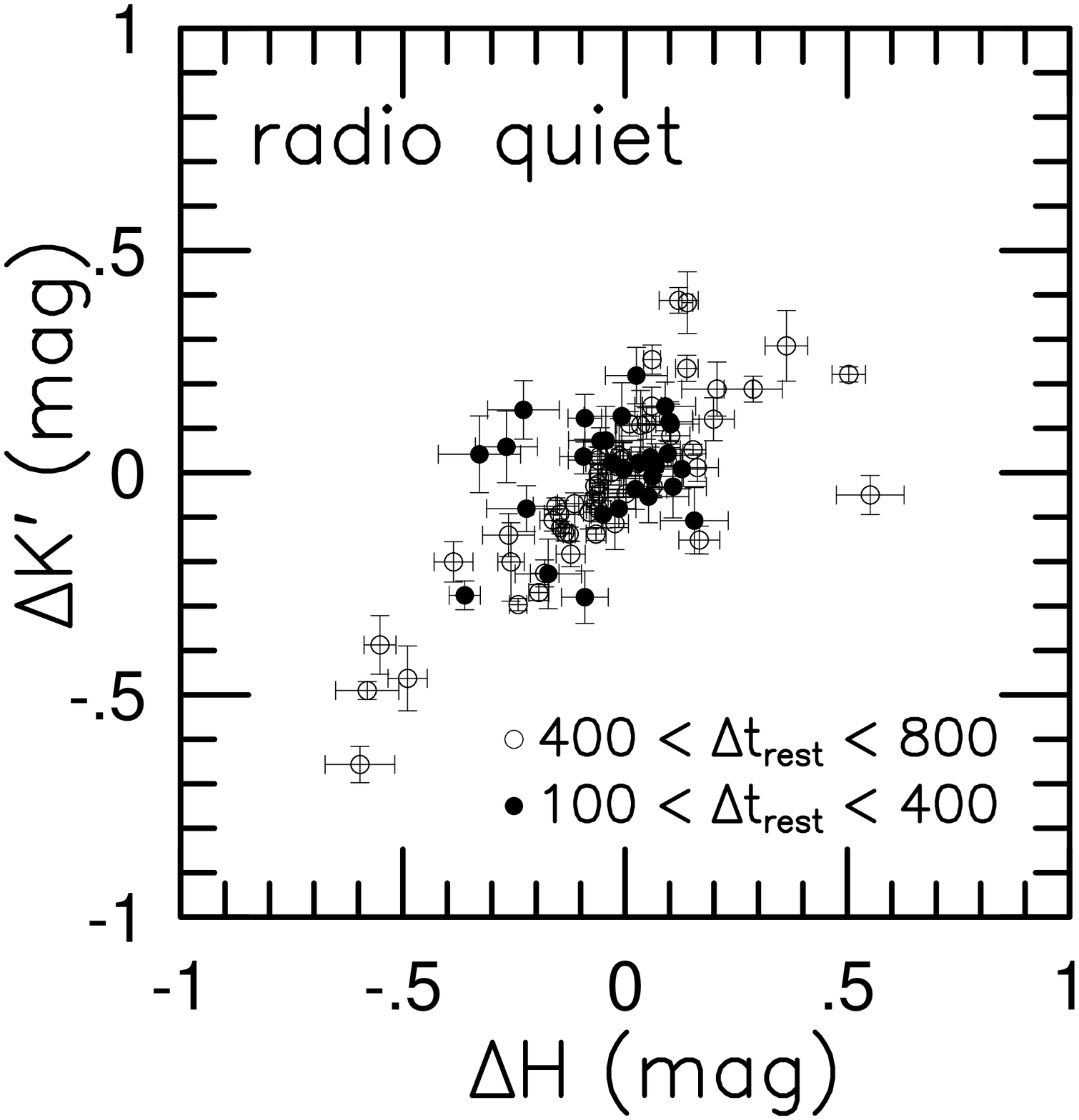}
\end{figure}

\clearpage
\begin{figure}
  Figure 7\\
  \epsscale{0.5}
  \plotone{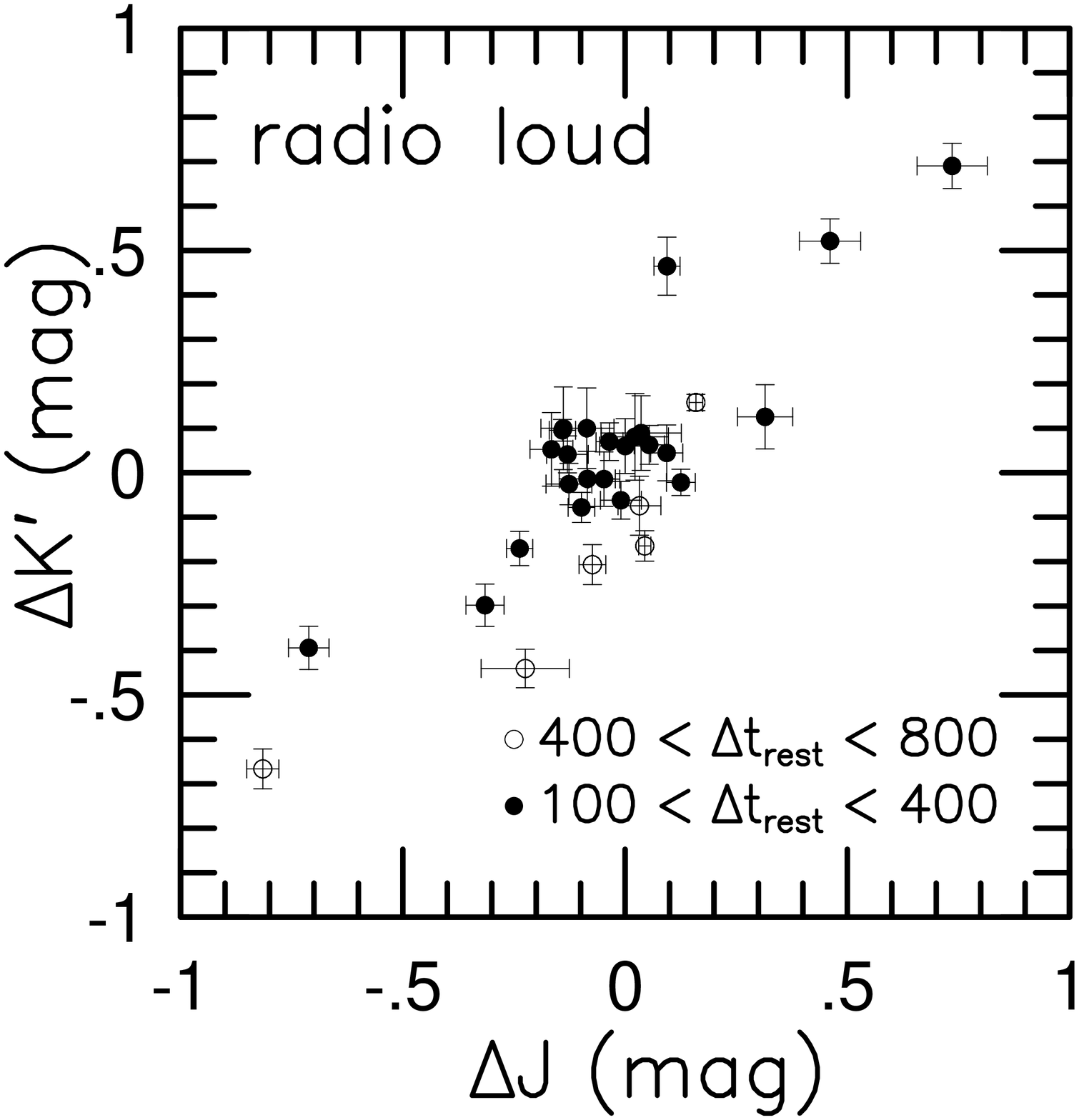}
\\
  \plotone{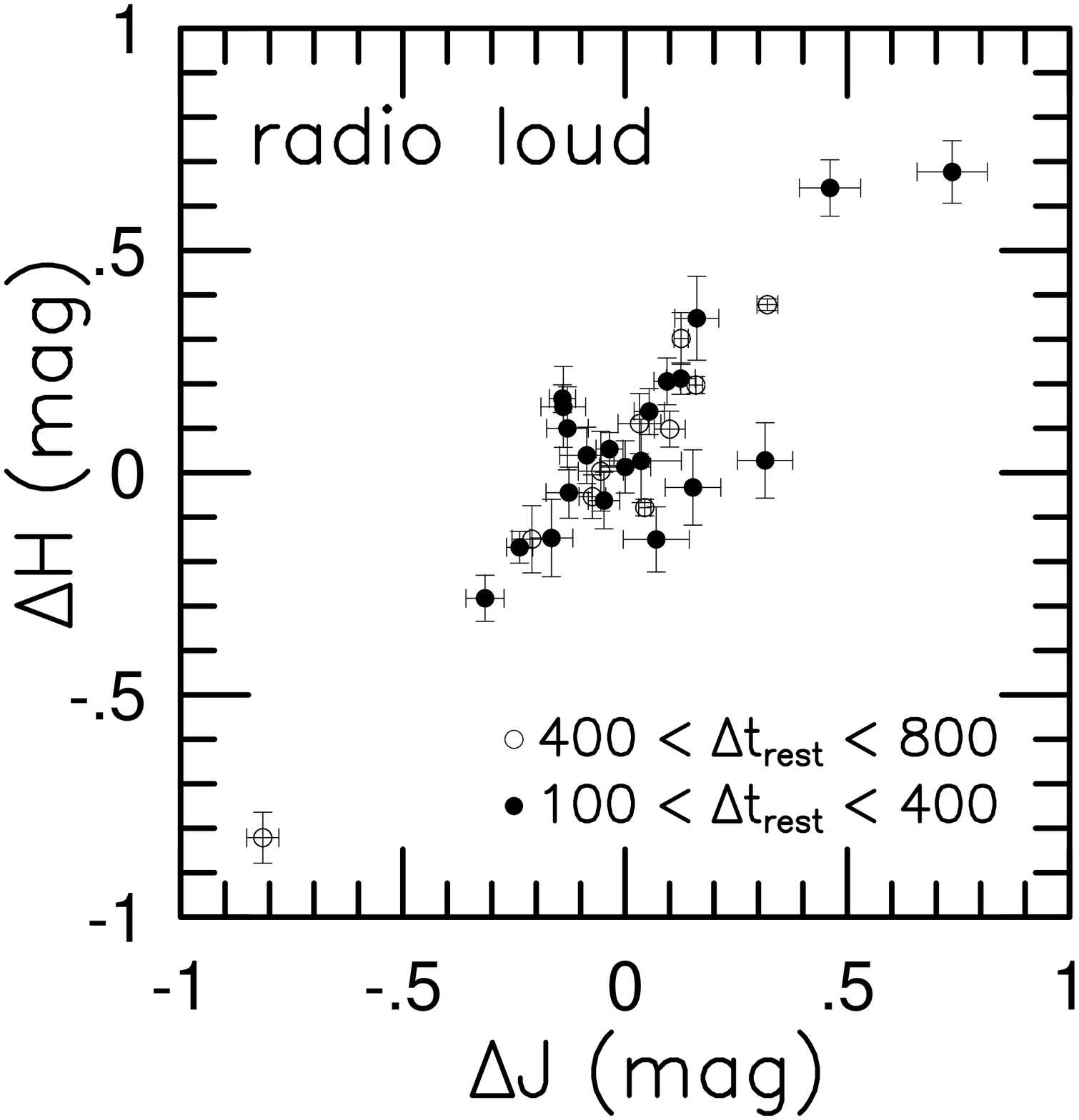}
  \plotone{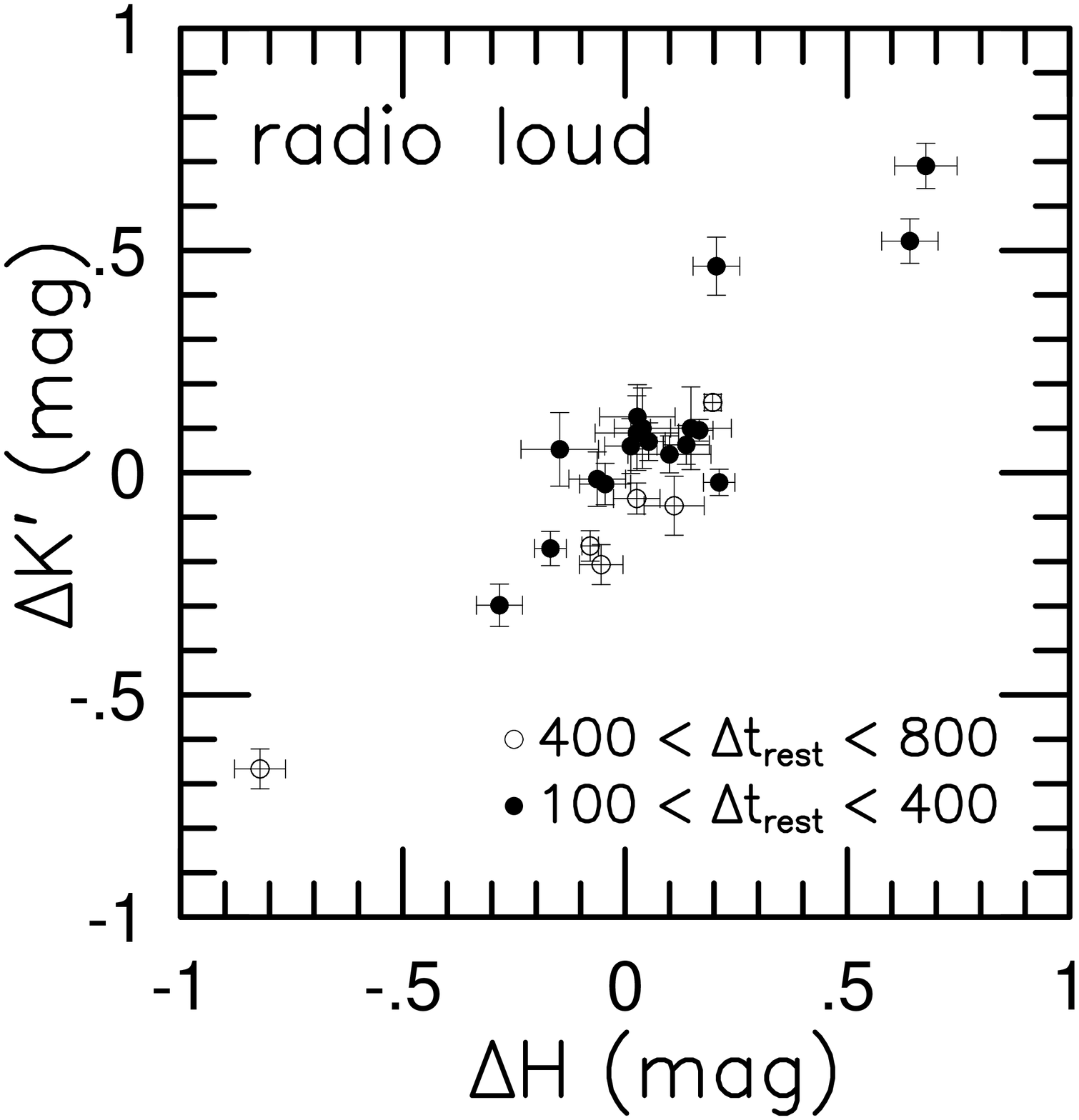}
\end{figure}

\clearpage
\begin{figure}
  Figure 8\\
  \epsscale{1}
  \plotone{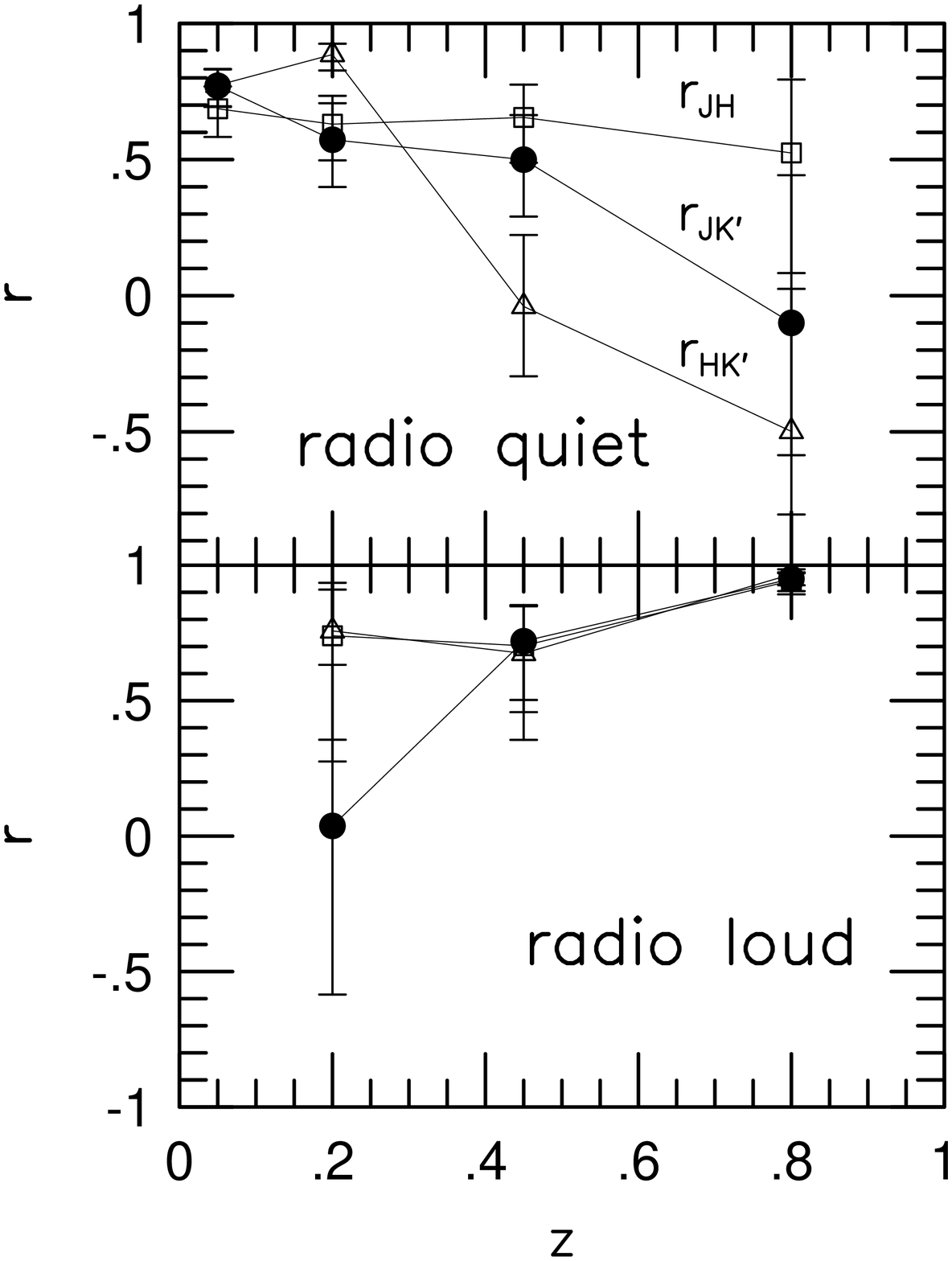}
\end{figure}

\clearpage
\begin{figure}
  Figure 9\\
  \epsscale{1}
  \plotone{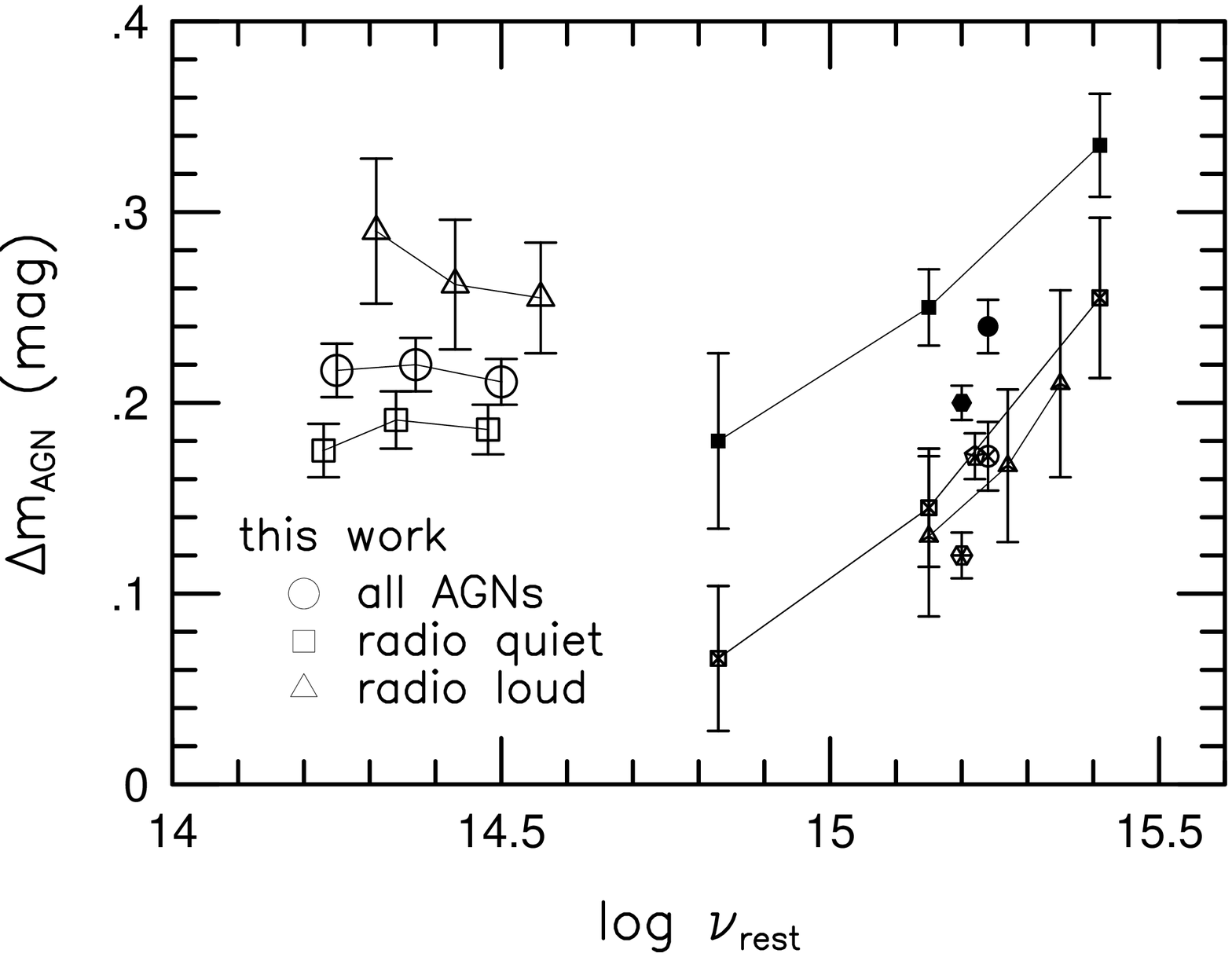}
\end{figure}

\clearpage
\begin{figure}
  Figure 10\\
  \epsscale{1}
  \plotone{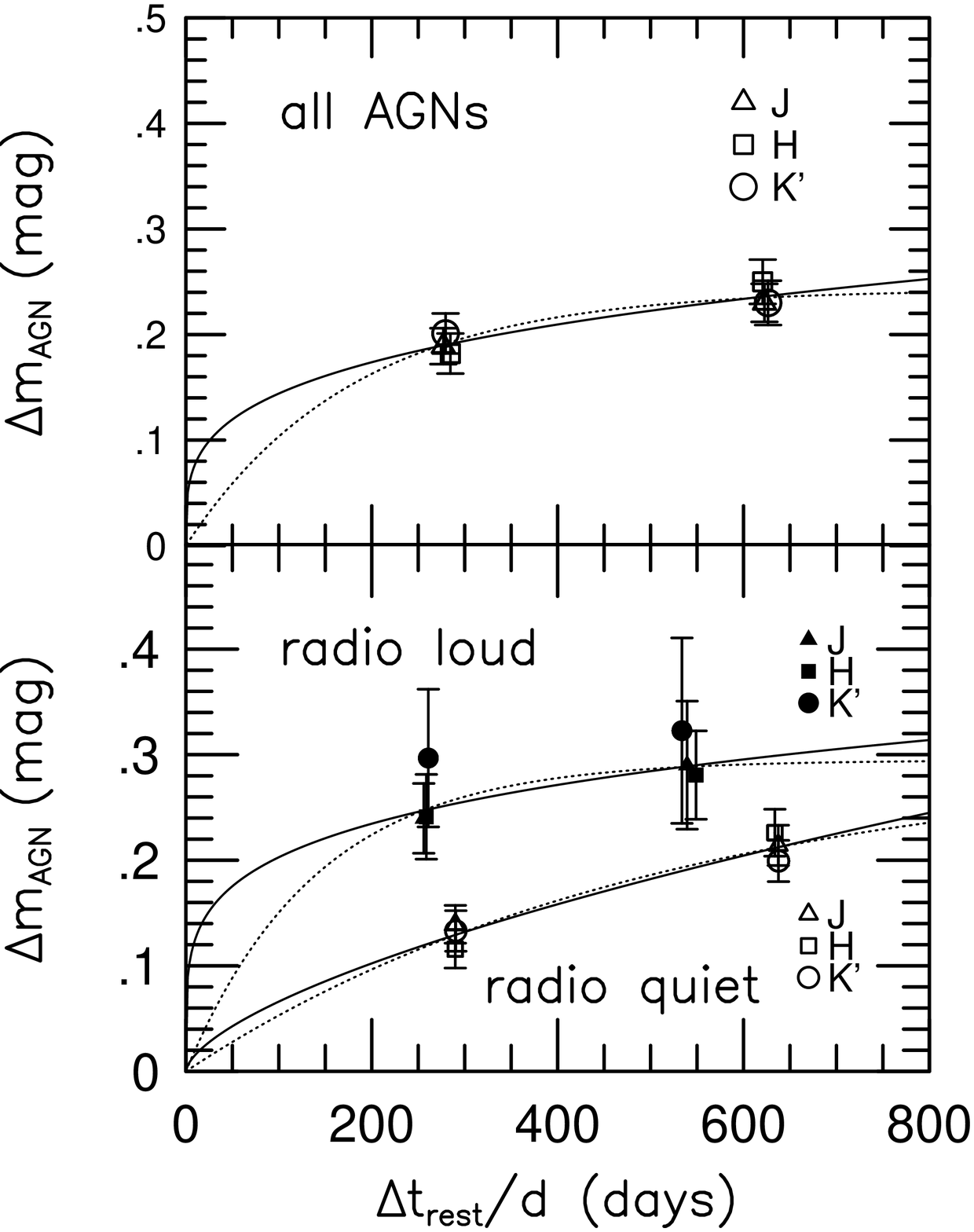}
\end{figure}

\clearpage
\begin{figure}
  Figure 11\\
  \epsscale{1}
  \plotone{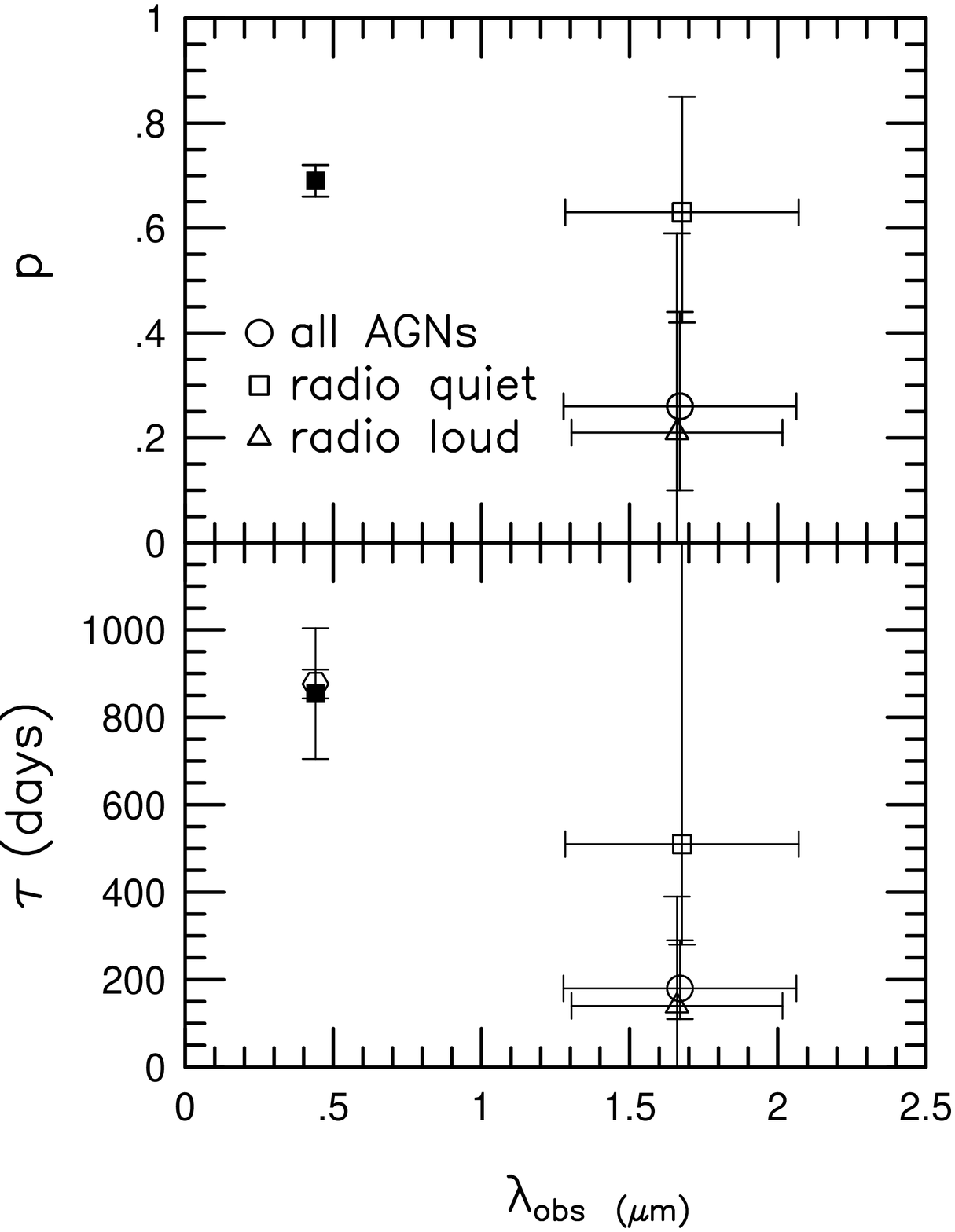}
\end{figure}

\clearpage
\begin{figure}
  Figure 12\\
  \epsscale{1}
  \plotone{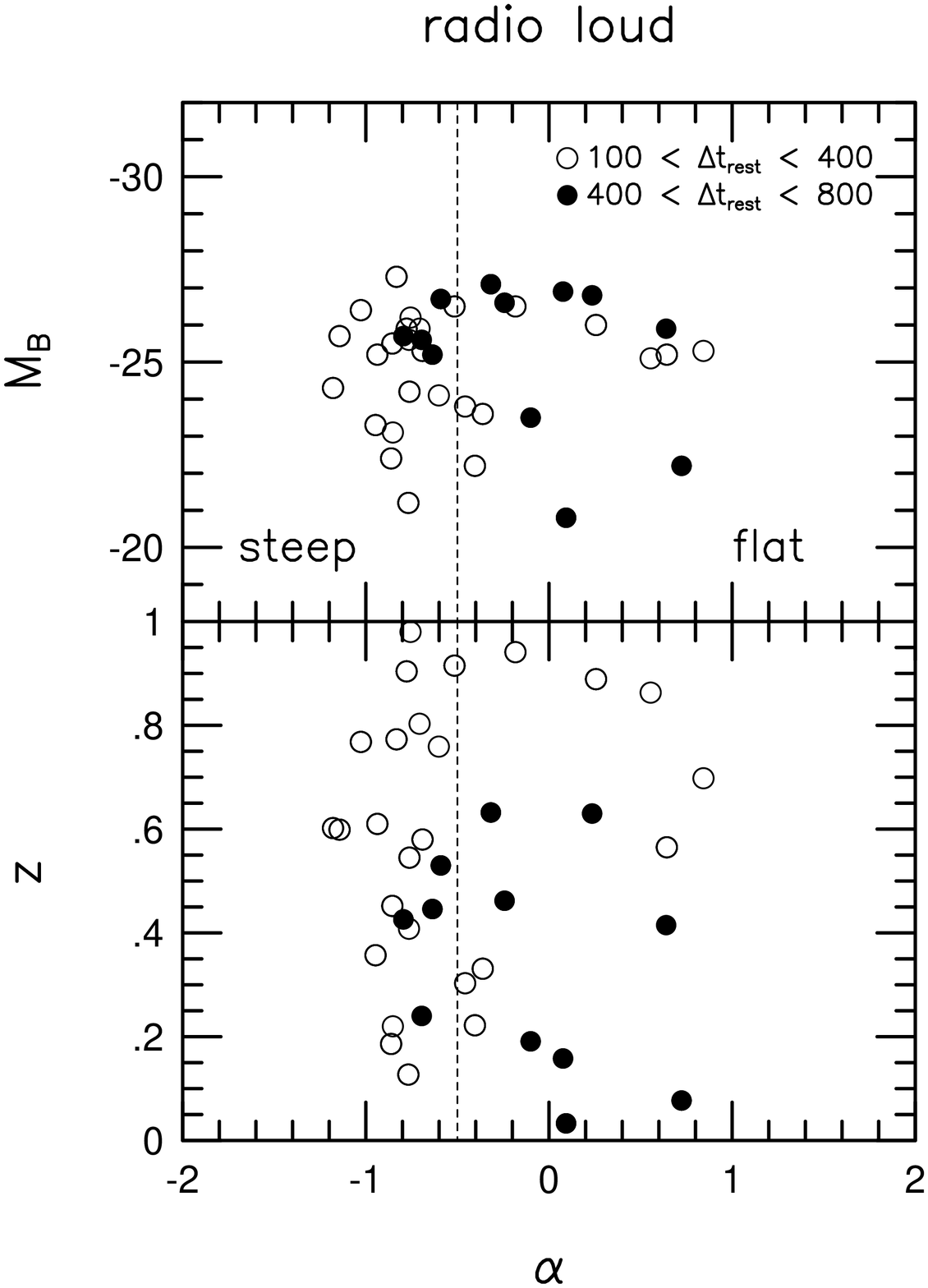}
\end{figure}

\clearpage
\begin{figure}
  Figure 13\\
  \epsscale{1}
  \plotone{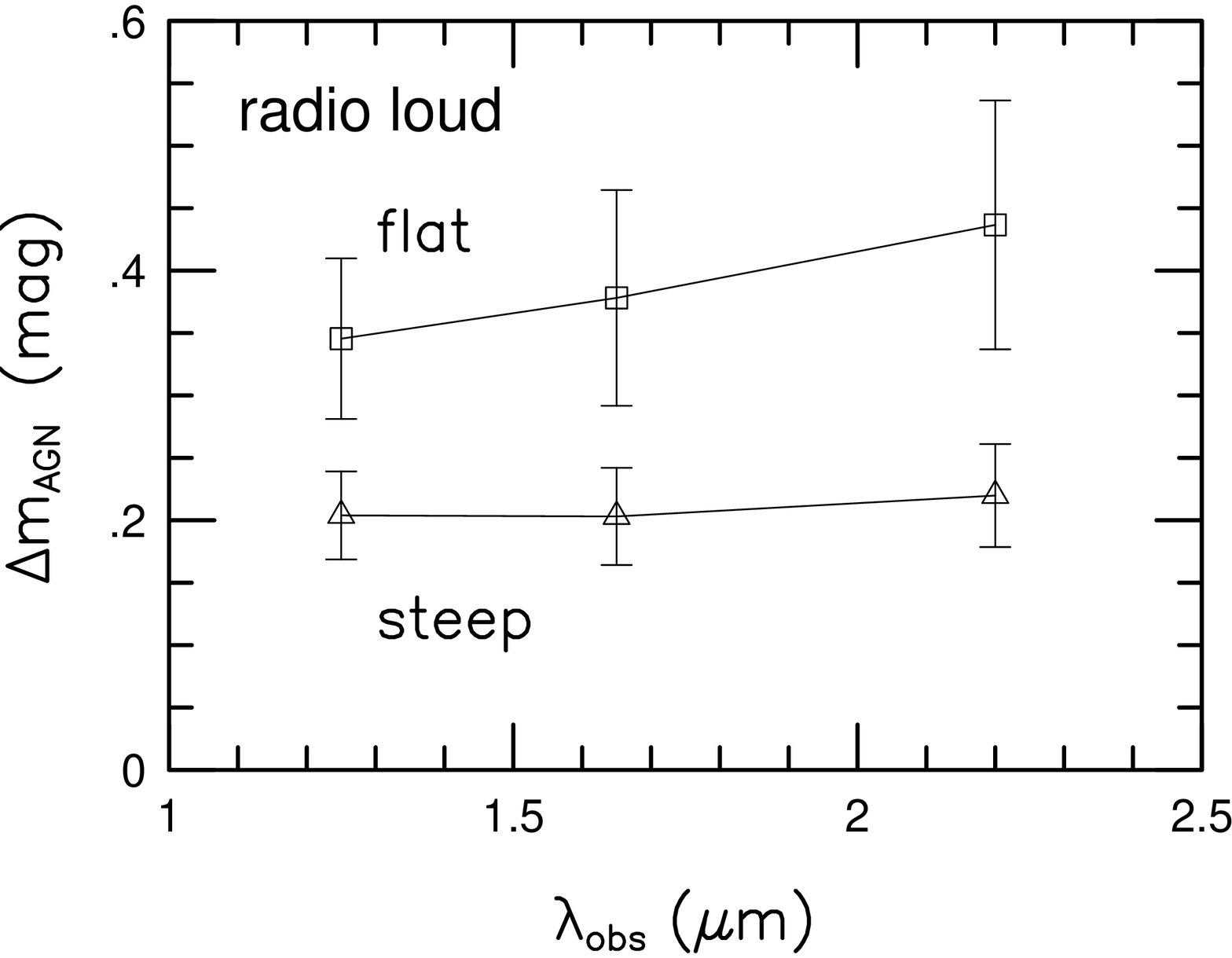}
\end{figure}


\clearpage

\begin{deluxetable}{lccccc}
\footnotesize
\tablenum{1}
\tablewidth{0pt}
\tablecaption{Ratio of ensemble variabilities in two groups of AGNs}
\tablehead{
 & \multicolumn{4}{c}{$\Delta m(``a")/\Delta m(``b")$ } 
  \nl \cline{2-5}
  \colhead{Tested property}& 
  \colhead{$J$} &
  \colhead{$H$ } & 
  \colhead{$K'$ } &
  \colhead{Average} &
}
\startdata
Radio strength .......... & 1.37  & 1.37  & 1.65  & 1.46 & \nl  
$\Delta t_{\rm rest}$ ........................ & 1.22  & 1.37  & 1.14 & 
1.24 & \nl    
$M_B$ ........................... & 1.29  & 1.18  & 1.14 & 1.20 & \nl    
Redshift .................... & 1.05  & 0.95  & 1.16 & 1.05  & \nl   
Seyfert type .............. & 1.06  & 1.04  & 0.92 & 1.01 & \nl   
$J$-$H$ .......................... & 1.02  & 1.00  & 1.01 & 1.01  & \nl   
$H$-$K'$ ........................ & 1.25  & 1.07  & 0.99 & 1.10 &
\enddata
\tablecomments{The entire sample is divided by each parameter into the
$``a"$ and $``b"$ groups, such as  
($f_{\nu}({\rm 6cm})/f_{\nu}(V)>100$, 
$f_{\nu}({\rm 6cm})/f_{\nu}(V)<10$), 
($400<\Delta t_{\rm rest}<800$, $100<\Delta t_{\rm rest}<400$),
($M_B<-23.5$, $M_B>-23.5$), ($0<z<0.3$, $z>0.3$), 
(Seyfert $1-1.5$, Seyfert $1.8-2$), ($J-H<0.8$, $J-H>0.8$), and
($H-K'<0.8$, $H-K'>0.8$).
}
\end{deluxetable}

\clearpage
\begin{deluxetable}{lcccccc}
\footnotesize
\tablenum{2}
\tablewidth{0pt}
\tablecaption{Test for wavelength dependence of the ensemble 
variability}
\tablehead{
 & & & & \multicolumn{2}{c}{Confidence interval of $a_2$} \nl \cline{5-6} 
 \multicolumn{1}{c}{Sample} & \colhead{ $a_1$} & \colhead{ $a_2$} 
 & & \colhead{95\%} & \colhead{99\%} &
}
\startdata
All AGNs  ......................  & 0.205 & $\;$0.029  && -0.192 $-$ 0.242  
& -0.248 $-$ 0.295 & \nl        
Radio-quiet AGNs ........ & 0.205  & -0.065  && -0.333 $-$ 0.193  
& -0.403 $-$ 0.258 & \nl
\hspace{5mm} Short $\Delta t_{\rm rest}$ ............ & 0.139  & -0.044  && 
-0.575 $-$ 0.452 
& -0.741 $-$ 0.593 & \nl
\hspace{5mm} Long $\Delta t_{\rm rest}$ ............ & 0.242  & -0.079  && 
-0.404 $-$ 0.234 
& -0.494 $-$ 0.316 & \nl
Radio-loud AGNs ......... & 0.213  & $\;$0.136  && -0.342 $-$ 0.582 
& -0.470 $-$ 0.693 & \nl
\hspace{5mm} Short $\Delta t_{\rm rest}$ ............ & 0.190  & $\;$0.169  
&& -0.398 $-$ 0.703 
& -0.556 $-$ 0.841 & \nl
\hspace{5mm} Long $\Delta t_{\rm rest}$ ............. & 0.251  & $\;$0.111  
&& -0.952 $-$ 0.992 
& -1.313 $-$ 1.228 &
\enddata
\tablecomments{
    The test was done by adopting a two-parameter function of 
$\Delta m_{\lambda}=a_1\exp(a_2\lambda)\label{eq_v_fit}$.
    }
\end{deluxetable}

\clearpage
\begin{deluxetable}{llccrr}
\footnotesize
\tablenum{3}
\tablewidth{0pt}
\tablecaption{Test for radio strength, $\Delta t_{\rm rest}$, $M_B$, 
  and $z$ dependence of the ensemble variability }
\tablehead{
  \colhead{Sample}  &  Tested property  & \colhead{ $a_1$}   &
\colhead{$\sigma_{a_1}$}  & \multicolumn{1}{c}{$\chi^2$}   &
\multicolumn{1}{c}{$P$(\%)}  
}
\startdata
All AGNs ......................  & Radio strength  & 0.196 & 0.007 & 15.767 
& $>$99.9 \nl
Radio-quiet AGNs ........ & $\Delta t_{\rm rest}$ & 0.167 & 0.008 & 27.797 
& $>$99.9 \nl
\hspace{5mm} Short $\Delta t_{\rm rest}$ ............ & $M_B$ & 0.126 & 
0.011 & 2.757 & 75.0 \nl
   \hspace{25mm} ............ & z    & 0.115 & 0.010 & 6.179 & 95.5 \nl
\hspace{5mm} Long $\Delta t_{\rm rest}$ ............ & $M_B$ & 0.212 & 
0.012 & 1.167 & 44.0 \nl
   \hspace{25mm} ............ & z    & 0.196 & 0.011 & 14.692 & $>$99.9 \nl
Radio-loud AGNs .......... & $\Delta t_{\rm rest}$ & 0.262 & 0.019 & 1.125 
& 71.1 \nl
\hspace{5mm} Short $\Delta t_{\rm rest}$ ............ & $M_B$ & 0.131 & 
0.016 & 35.898 & $>$99.9 \nl
   \hspace{25mm} ............ & z     & 0.117 & 0.013 & 39.503 & $>$99.9 \nl
\enddata 
\tablecomments{
  $P$ represents the reliability of rejecting the hypothesis that
  $\Delta m_{\lambda}=a_1$ does not depend on the tested property.
}
\end{deluxetable}


\begin{deluxetable}{lcccc}
\footnotesize
\tablenum{4}
\tablewidth{0pt}
\tablecaption{Correlation coefficient $r_{ij}$ of variabilities in the
$\lambda_i$ and $\lambda_j$ bands}
\tablehead{
  \multicolumn{1}{c}{Sample} & \colhead{$\lambda_i$, $\lambda_j$} 
  & $n$ & \colhead{$r_{ij}$} & Confidence interval (68.3\%) 
}
\startdata
All AGNs ......................   & $J$, $H$   & 127  & 0.74  & 0.70 $-$ 0.78 \nl
 \hspace{21mm} ................ & $H$, $K'$  & 111  & 0.81  & 0.77 $-$ 0.84 \nl
 \hspace{21mm} ................ & $J$, $K'$  & 118  & 0.71  & 0.66 $-$ 0.76 \nl
Radio-quiet AGNs .........   & $J$, $H$   &  91  & 0.65  & 0.59 $-$ 0.71 \nl
 \hspace{21mm} ................ & $H$, $K'$  &  84  & 0.72  & 0.66 $-$ 0.77 \nl
 \hspace{21mm} ................ & $J$, $K'$  &  85  & 0.65  & 0.59 $-$ 0.71 \nl
Radio-loud AGNs ..........      & $J$, $H$   &  31  & 0.88  & 0.83 $-$ 0.91 \nl
 \hspace{21mm} ................ & $H$, $K'$  &  24  & 0.91  & 0.87 $-$ 0.94 \nl
 \hspace{21mm} ................ & $J$, $K'$  &  30  & 0.86  & 0.80 $-$ 0.90  
\enddata
\tablecomments{ 
     $n$ represents the number of AGNs in a sample. 
     }
\end{deluxetable}

\clearpage
\begin{deluxetable}{lccccc}
\footnotesize
\tablenum{5}
\tablewidth{0pt}
\tablecaption{Equivalence test for the correlation 
coefficients $r_{JH}$, $r_{HK'}$, and $r_{JK'}$ }
\tablehead{
 \multicolumn{1}{c}{Sample}   
& \colhead{$r_{JH}$}   & \colhead{$r_{HK'}$}  
& \colhead{$r_{JK'}$}   & \colhead{$r_{\rm true}$}  & \colhead{$P$(\%)}
}
\startdata
Radio-quiet AGNs &  &  &  &  &  \nl
 \hspace{5mm}  $z=0.0-0.1$ ...... & 0.69 (36) & $\;$0.77 (38) & $\;$0.77 
(38) & 0.75   & 31.5  \nl
\hspace{5mm}  $z=0.1-0.3$ ...... & 0.63 (29) & $\;$0.89 (23) & $\;$0.57 (22) & 
\hspace{3mm}0.72$^1$ & 96.8  \nl
\hspace{5mm}  $z=0.3-0.6$ ...... & 0.65 (19) & -0.04 (17) & $\;$0.50 (19) & 
0.42  & 58.1  \nl
\hspace{5mm}  $z=0.6-1.0$ ...... & 0.52  (7) & -0.51  (6) & -0.11  (6) & 
0.03  & 68.3  \nl
Radio-loud AGNs &  &  &  &  &  \nl
\hspace{5mm}  $z=0.1-0.3$ ...... & 0.74  (6)  & $\;$0.76  (5) & $\;$0.04  
(5) & 0.60  & 46.0 \nl
\hspace{5mm}  $z=0.3-0.6$ ...... & 0.70 (10)  & $\;$0.67  (8) & $\;$0.72 
(11) & 0.70  & 1.17 \nl 
\hspace{5mm}  $z=0.6-1.0$ ...... & 0.94 (12)  & $\;$0.97  (9) & $\;$0.95 
(12) & 0.95  & 12.9 
\tablecomments{ The number in the parensatheses is the number of AGNs in a 
sample. $r_{\rm true}$ 
 is the estimation of true correlation coefficient.
 $P$ is the reliability of rejecting the hypothesis that $r_{JH}$, 
$r_{HK'}$, and $r_{JK'}$ 
 are equivalent to each other.  \nl
 1: The value of $r_{\rm true}=0.72$ is not adequate since the equivalence 
is rejected.
        }
\enddata
\end{deluxetable}


\clearpage
\begin{deluxetable}{llcccc}
\footnotesize
\tablenum{6a}
\tablewidth{0pt}
\tablecaption{Fitted values of the parameters in a function of   
$\Delta m=A(\Delta t_{\rm rest})^p$}
\tablehead{
 & & & \multicolumn{2}{c}{Confidence interval of $p$} & \nl \cline{4-5}
 \multicolumn{1}{c}{Sample} & \multicolumn{1}{c}{$A$} 
 & \colhead{$p$} & \colhead{95\%} & \colhead{99\%} & 
}
\startdata
All AGNs ...................... & 0.044  & 0.26  & \hspace{9mm}$-$ 0.53 & 
\hspace{9mm}$-$ 0.60 \nl
Radio-quiet AGNs .........      & 0.0036 & 0.63  & 0.30 $-$ 1.00  & 0.22 
$-$ 1.09 \nl
Radio-loud AGNs ..........      & 0.077  & 0.21  & \hspace{9mm}$-$ 0.79 & 
\hspace{10mm}$-$ 0.91
\tablecomments{ The cases with no lower limit to the confidence interval 
are those for which the
lower limit becomes negative. 
        }
\enddata
\end{deluxetable}

\clearpage
\begin{deluxetable}{llcccc}
\footnotesize
\tablenum{6b}
\tablewidth{0pt}
\tablecaption{Fitted values of the parameters in a function of   
$\Delta m=B(1- \exp (-\Delta t_{\rm rest}/\tau))$}
\tablehead{
 & & & \multicolumn{2}{c}{Confidence interval of $\tau$} & \nl \cline{4-5}
 \multicolumn{1}{c}{Sample} & \multicolumn{1}{c}{$B$} 
 & \colhead{$\tau$} & \colhead{95\%} & \colhead{99\%} & 
}
\startdata
All AGNs ...................... & 0.24 & 180  & 50 $-$ 370 & 
\hspace{5mm}$-$ 440 \nl
Radio-quiet AGNs .........  & 0.30 & 510  & 200 $-$\hspace{9mm} & 170 
$-$\hspace{9mm} \nl
Radio-loud AGNs ..........  & 0.29 & 140  & \hspace{5mm}$-$ 820 & 
\hspace{8mm}$-$ 2040
\tablecomments{ The cases with no lower limit to the confidence interval 
are those for which the 
lower limit becomes negative. The cases with no upper limit are those for 
which linear 
relation of $\Delta m\propto \Delta t_{\rm rest}$ is accepted from the 
$\chi^2$ test.       }
\enddata
\end{deluxetable}

\clearpage
\begin{deluxetable}{lcccc}
\footnotesize
\tablenum{7}
\tablewidth{0pt}
\tablecaption{AGNs in the literature which support a dust 
 reverberation model. }
\tablehead{
\colhead{Object} & \colhead{\rm Ref} & \colhead{$z$}   &
\colhead{$M_B$}  & 
\colhead{$f_\nu({\rm 6cm})/f_\nu(V)$ }
}
\startdata
 GQ COM  ........... & 1 & 0.165  & -24.4  & 0.0  \nl
 Fairall 9 ............. & 2 & 0.046  & -23.0  & 0.0  \nl    
 NGC 3783 ......... & 3 & 0.009  & -19.7  & 0.7  \nl    
 MARK 744 ........ & 4 & 0.010  & -19.3  & 0.0  \nl   
 NGC 4151 ......... & 5 & 0.003  & -18.7  & 2.3  \nl  
 NGC 1566 ......... & 6 & 0.004  & -18.0  & 3.7
\enddata  
\tablecomments{ The data of $z$, $M_B$, 
$f_\nu ({\rm 6cm})$, and $f_\nu (V)$ 
are taken from the VV catalog.  
All AGNs listed here are regarded as radio quiet satisfying
our criterion of $f_\nu ({\rm 6cm})/f_\nu (V)<10$. \\
References:  
(1) Sitko, M. L. et al. 1993,  
(2) Clavel, J., Wamstecker, W. and Glass, I. S. 1989,
(3) Glass, I. S. 1992,
(4) Nelson, B. O. 1996b,
(5) Oknyanskii, V. L. 1993,
(6) Baribaud, T. et al. 1992.
}
\end{deluxetable}

\clearpage
\begin{deluxetable}{lccc}
\footnotesize
\tablenum{8}
\tablewidth{0pt}
\tablecaption{Statistics of AGNs based on Nelson's 
$V$ and $K$ band monitoring observations. }
\tablehead{
 & \multicolumn{3}{c}{Number of AGNs in a sample} \nl \cline{2-4}
 \multicolumn{1}{c}{Sample}  & \colhead{Radio qiet}   & \colhead{Radio loud}  
 & \colhead{Ambiguous}
}
\startdata
All AGNs ............................................ & 43 & 5 & 3 \nl
AGNs with detected variability$^1$ ........  & 29 & 3 & 1 \nl
AGNs with time delay obtained$^2$  .......  &  6 & 0 & 0 
\tablecomments{ The result of  variability and time delay is 
  taken from Nelson (1996a). \nl 
 1: AGNs for which the variability is detected in either $V$ or $K$ band or 
in both. \nl
 2: AGNs for which the variability is detected in both $V$ and $K$ bands 
and the time delay 
    of light variation in the $K$ band relative to the $V$ band is measured.  
}
\enddata
\end{deluxetable}

\end{document}